\documentclass[opre,nonblindrev]{informs3_hide}

\OneAndAHalfSpacedXI % current default line spacing
\usepackage{endnotes}
\let\footnote=\endnote

\usepackage{bbm,multirow,multicol,enumitem} % general packages
\usepackage[ruled,linesnumbered,algo2e,norelsize]{algorithm2e}
\usepackage[dvipsnames]{xcolor} % color
\usepackage{tikz} % tikz
\usetikzlibrary{arrows}
\usepackage{pifont}

\newcommand{\bE}{\mathbb{E}}
\newcommand{\bR}{\mathbb{R}}
\newcommand{\bI}{\mathbbm{1}}
\newcommand{\ve}{\varepsilon}

\newcommand{\OPT}{\mathsf{OPT}}

\newcommand{\OPTLP}{\mathsf{OPT}_{\mathsf{LP}}}
\newcommand{\ALG}{\mathsf{ALG}}

\newcommand{\VK}{\mathsf{A1}}
\newcommand{\VKp}{\mathsf{A1'}}
\newcommand{\Exp}{\mathsf{Exp}}
\newcommand{\Samp}{\mathsf{Samp}}
\newcommand{\CR}{c^*}

\newcommand{\pvk}{P^{\VK}}
\newcommand{\pvkp}{P^{\VKp}}
\newcommand{\pexp}{P^{\Exp}}
\newcommand{\psamp}{P^{\Samp}}

\newcommand{\fexp}{F^{\Exp}}
\newcommand{\fsamp}{F^{\Samp}}

\newcommand{\vexp}{V^{\Exp}}
\newcommand{\vsamp}{V^{\Samp}}

\newcommand{\level}{\mathtt{level}}
\newcommand{\val}{\mathtt{val}}
\newcommand{\sold}{\mathtt{sold}}
\newcommand{\true}{\mathtt{true}}
\newcommand{\false}{\mathtt{false}}
\newcommand{\qtot}{\mathtt{q\_tot}}
\newcommand{\inventory}{\mathtt{inventory}}

\newcommand{\cA}{\mathcal{A}}
\newcommand{\cF}{\mathcal{F}}
\newcommand{\cG}{\mathcal{G}}
\newcommand{\cP}{\mathcal{P}}
\newcommand{\cS}{\mathcal{S}}
\newcommand{\vg}{\mathbf{g}}
\newcommand{\vG}{\mathbf{G}}
\newcommand{\vh}{\mathbf{h}}
\newcommand{\vH}{\mathbf{H}}
\newcommand{\vu}{\mathbf{v}}

\newcommand{\rmin}{r^{\min}}
\newcommand{\rmax}{r^{\max}}

\newcommand{\legendLength}{0.2}
\newcommand{\beforeAfterShift}{10}

\usepackage{natbib}
\bibpunct[, ]{(}{)}{,}{a}{}{,}%
\def\bibfont{\small}%
\TheoremsNumberedThrough     % Preferred (Theorem 1, Lemma 1, Theorem 2)
\ECRepeatTheorems

\EquationsNumberedThrough    % Default: (1), (2), ...
\MANUSCRIPTNO{OPRE-2018-10-607.R1} % When the article is logged in and DOI assigned to it,
\begin{document}
%%%%%%%%%%%%%%%%

%\setcounter{page}{1}
%\DOI{10.1287/xxxx.0000.0000}%

\RUNAUTHOR{Ma, Simchi-Levi, and Teo}

\RUNTITLE{Single-leg RM with Limited Demand Information}

% \TITLE{Bundling Information Goods of Decreasing Value}
\TITLE{On Policies for Single-leg Revenue Management with Limited Demand Information}

\ARTICLEAUTHORS{%
\AUTHOR{Will Ma}
\AFF{Graduate School of Business, Columbia University, New York, NY 10027, \EMAIL{wm2428@gsb.columbia.edu}} %, \URL{}}
\AUTHOR{David Simchi-Levi}
\AFF{Institute for Data, Systems, and Society, Department of Civil and Environmental Engineering, and Operations Research Center, Massachusetts Institute of Technology, Cambridge, MA 02139, \EMAIL{dslevi@mit.edu}}
\AUTHOR{Chung-Piaw Teo}
\AFF{NUS Business School, National University of Singapore, Singapore, \EMAIL{bizteocp@nus.edu.sg}}
}
\ABSTRACT{

In this paper we study the single-item revenue management problem, with no information given about the demand trajectory over time.
When the item is sold through accepting/rejecting different fare classes, Ball and Queyranne (2009) have established the tight competitive ratio for this problem using booking limit policies, which raise the acceptance threshold as the remaining inventory dwindles.
However, when the item is sold through dynamic pricing instead, there is the additional challenge that offering a low price may entice high-paying customers to substitute down.
We show that despite this challenge, the same competitive ratio can still be achieved using a randomized dynamic pricing policy.
Our policy incorporates the price-skimming technique from Eren and Maglaras (2010), but importantly we show how the randomized price distribution should be stochastically-increased as the remaining inventory dwindles.
A key technical ingredient in our policy is a new ``valuation tracking'' subroutine, which tracks the possible values for the optimum, and follows the most ``inventory-conservative'' control which maintains the desired competitive ratio.
Finally, we demonstrate the empirical effectiveness of our policy in simulations, where its average-case performance surpasses all naive modifications of the existing policies.

}

\KEYWORDS{online algorithms, competitive ratio, revenue management, dynamic pricing}

%\HISTORY{This paper was
%first submitted on April 12, 1922 and has been with the authors for
%83 years for 65 revisions.}

\maketitle
%%%%%%%%%%%%%%%%%%%%%%%%%%%%%%%%%%%%%%%%%%%%%%%%%%%%%%%%%%%%%%%%%%%%%%

\section{Introduction}

In the single-leg revenue management problem, a firm is selling multiple products that share a single capacity over a finite time horizon.
The prices of the products and the unreplenishable starting capacity are all exogenously determined.
The firm's objective is to maximize its total revenue earned, by dynamically controlling the availability of different products over time.
The tradeoff lies between ``myopic'' controls which maximize revenue at an immediate point in time, and ``conservative'' controls which preserve capacity for the remaining time horizon.

This problem was originally motivated by airlines.
Indeed, for a single flight leg, it has a limited seat capacity, and the products correspond to different ``fare classes'' (e.g.\ economy, basic economy) dictating prices at which seats could be sold.
The flight's seat capacity and fare classes have been determined long in advance, through factors such as business strategy and competitor price points.
Finally, the time horizon is finite, ending upon the flight's departure.

We study this problem in the setting where very limited information is given or can be learned about demand.
This setting was introduced by \citet{BQ09,LGBK08}, who derive heuristic controls based on only knowledge of the fare class prices.
These papers consider \textit{booking limit} policies, which can be described as follows.
Initially, all of the fare classes are made available to customers.
Once the fraction of total seats sold surpasses a critical threshold, the lowest fare becomes unavailable.
Progressively higher fares are made unavailable until the flight either becomes full or takes off.

An important assumption made in these papers is that the demands for the different fare classes are \textit{independent}.
That is, although the lower fares are made available until their booking thresholds are reached, there is no risk of \textit{cannibalization} in sales, where a customer who would have accepted a higher fare \textit{substitutes} down into buying a lower fare.
A classical justification for this is that the fare classes are designed to segment customers and achieve price discrimination---i.e., the drawbacks of basic economy are supposed to dissuade price-insensitive travelers from buying down.

However, fare classes rarely achieve perfect price discrimination, and in the age of e-commerce, increasingly many airlines are introducing overlapping fare families despite the cannibalization \citep{fiig2012forecasting}.
Furthermore, most items (unlike seats on an airplane) are sold using a single price and not packaged into fare classes at all.
Motivated by these two facts, our goal is to derive advanced pricing controls for the single-leg revenue management problem under limited demand information, which do not require the assumption of no cannibalization.

\subsection{Dynamic Pricing Model}\label{sec::setupIntro}

In this paper, we focus on the dynamic pricing problem of a single retail item being marked at a single price, with no fare classes at all.
To connect back to airlines, we can interpret this as offering an assortment with all fare classes greater than or equal to the posted price.
In the conclusion of the paper, we explain why this immediately implies the same results for the airline model with fare classes and substitution, as long as substitution follow a random-utility choice model.

We let $\cP$ denote the set of prices which the firm could feasibly offer at any point in time,
corresponding to the set of fare classes discussed earlier.
We usually assume $\cP$ to be discrete for ease of exposition and computation.
In \textbf{Section~\ref{sec::continuum}}, we allow for $\cP$ to be any continuous price interval, and derive comparable theoretical results to those when the interval is approximated using discrete prices.
Discreteness of $\cP$ is further justified in industries with established price points to charge for items, e.g.\ prices ending in 99 cents.

The dynamic pricing problem proceeds as follows.
The item starts with $k$ units of discrete \textit{inventory}.
Sequentially-arriving customers $t=1,\ldots,T$ each have a \textit{valuation} $V_t$, denoting the maximum price in $\cP$ that the customer is willing to pay.
We colloquially refer to customer $t$ as arriving in ``time $t$'';
%and allow $V_t$ to be 0 to represent a ``dummy'' customer with no interest in the item.
however, we emphasize that $t$ does not correspond to an actual unit of time and that the total number of customers $T$ is unknown.
An \textit{online algorithm} sequentially charges a price $P_t$ for each customer $t$, and if $V_t\ge P_t$, then revenue $P_t$ is earned and one unit of inventory is depleted.

In settings where no information is given about demand, an online algorithm is evaluated by comparing its total revenue earned on different sequences of valuations to that of a clairvoyant optimum.
For any such sequence $V_1,\ldots,V_T$, the \textit{offline optimum} $\OPT(V_1,\ldots,V_T)$ is defined as the maximum revenue that could have been earned from deterministically knowing all the valuations in advance, equal to the sum of the $k$ largest values in $V_1,\ldots,V_T$.
For $c\le1$, if an online algorithm can guarantee that its revenue is at least $c\cdot\OPT(V_1,\ldots,V_T)$ on every sequence $V_1,\ldots,V_T$, then it is said to be $c$-\textit{competitive}.
If $c$ is best-possible in that any (potentially randomized) online algorithm cannot uniformly guarantee better than $c\cdot\OPT(V_1,\ldots,V_T)$ revenue over all sequences $V_1,\ldots,V_T$, then $c$ is called the \textit{competitive ratio}.

We now describe the nature of $V_t$ and what is known about it before the online algorithm has to decide price $P_t$.
We consider the following variants.
\begin{enumerate}
\item \textbf{Public vs.\ Personalized Pricing}: In public pricing, $P_t$ must be decided before knowing anything about valuation $V_t$.
The motivation for this is an e-commerce platform that posts a public price without knowing which customer will arrive next.
After customer $t$ arrives and makes a purchase decision based on price $P_t$, (the distribution of) $V_t$ can be estimated based on her characteristics such as IP address (and this distribution can be posterior-updated based on her purchase decision).
The platform can then update its price $P_{t+1}$ for the next customer.

By contrast, in personalized pricing, (the distribution of) $V_t$ is given before having to decide $P_t$.
The motivation for this is an e-commerce platform that uses a customer's characteristics upon login to charge them a personalized price.  Like before, customer $t$ then makes a purchase decision based on price $P_t$.
However, in personalized pricing the online algorithm has more information (namely, about $V_t$) before having to make each decision $P_t$.
\item \textbf{Deterministic vs.\ Stochastic Valuations}: To illustrate our basic technique, and to relate to the model of \citet{BQ09}, we first assume that each valuation $V_t$ can be deterministically estimated based on the characteristics of customer $t$.
%, \textcolor{red}{after the price has been posted}.
We then extend to the setting where only a distribution for $V_t$ can be estimated each time.
We would like to point out in the public pricing setting,
clickstream data (the user's navigation path after they saw the price) can also be incorporated to generate these distributional estimates, which can greatly improve the accuracy as found by \citet{montgomery2004modeling}.
\end{enumerate}
We emphasize that under all combinations of variants, when deciding price $P_t$, (the distributions of) valuations $V_1,\ldots,V_{t-1}$ and the purchase decisions of customers $1,\ldots,t-1$ are known, and any information about valuations $V_{t+1},\ldots,V_T$ or the length $T$ is unknown.

\subsection{Outline of Theoretical Results}

Before stating our new results for the variants above, we first observe that---if \textit{personalized} pricing is permitted on \textit{deterministic} valuations, then our dynamic pricing model is equivalent to the multi-fare-class model of \citet{BQ09}.
Indeed, if the platform can always charge the maximum-willingness-to-pay of any customer it sells to, then it has achieved the same price discrimination assumed by fare classes.
\citet{BQ09} show in their model that the competitive ratio depends on only the price set $\cP$ and not the starting inventory $k$.
Letting $\CR(\cP)$ denote this optimal competitive ratio, the booking limit policies of \citet{BQ09} imply $\CR(\cP)$-competitive pricing policies in our model under the personalized, deterministic variant.

In light of this, our main theoretical contribution is the following results in progressively more difficult/general variants.
\begin{enumerate}
\item In \textbf{Section~\ref{sec::deterministic}}, we derive $\CR(\cP)$-competitive \textit{public} pricing policies for \textit{deterministic} valuations.
We motivate our main algorithmic idea in \textbf{Section~\ref{sec::attempts}}, illustrate it through geometrically ``stacking'' the valuations in \textbf{Sections~\ref{sec::stacking}}, and formalize it in \textbf{Section~\ref{sec::valuationTrackingProc}}.
We consider variations which satisfy intuitive properties that will become useful later, in \textbf{Sections~\ref{sec::modifiedAlg}--\ref{sec::structural}}.

\item In \textbf{Section~\ref{sec::stochastic}}, we extend our earlier ideas for deterministic valuations into $\CR(\cP)$-competitive public pricing policies for \textit{stochastic} valuations.
% in \textbf{Section~\ref{sec::exp_time}}.
A unique challenge arises in \textbf{Section~\ref{sec::poly_time}}, where we use sampling to convert a $\CR(\cP)$-competitive algorithm with exponential runtime into a $(\CR(\cP)-\ve)$-competitive algorithm with polynomial runtime.
\end{enumerate}
All of these policies are best-possible.  Indeed, since \citet{BQ09} show that a personalized pricing policy on deterministic valuations cannot be better than $\CR(\cP)$-competitive, any online policy in our more difficult/general variants cannot be better then $\CR(\cP)$-competitive.

Our policies use the \textit{price-skimming} technique of \citet{EM10}, who analyze how the price of an item should be distributed (e.g.\ across stores, across time) when the price set $\cP$ is known but demand is completely unknown.
Their model corresponds to ours when the inventory constraint is irrelevant (i.e.\ when $k\ge T$), in which case they show that the competitive ratio is also $\CR(\cP)$.
By contrast, our work shows how the price-skimming distribution should change over time based on remaining inventory.
%depend on inventory when it is relevant.  More specifically, our algorithm determines the optimal path-dependent pricing distribution for each time step $t$, conditional on the amount of remaining inventory.
In fact,
%in \textbf{Section~\ref{sec::structural}} we see that
our policies satisfy the following structural property:
\begin{quote}
``At any time step, the \textit{price distribution} which maximizes the \textit{competitive ratio} is \textit{strictly stochastically-decreasing} in the amount of remaining inventory.''
\end{quote}
This is analogous to a classical structural property when the demand sequence is distributionally-known in advance and dynamic programming can be used \citep{GvR94,ZZ00}:
\begin{quote}
``At any time step, the \textit{price} which maximizes the \textit{expected revenue} is \textit{strictly decreasing} in the amount of remaining inventory.''
\end{quote}
This structural property turns out to be crucial for good empirical performance.
In fact, in \textbf{Section~\ref{sec::rfpp}} we extend the results in \citet{EM10} to derive a $\CR(\cP)$-competitive public pricing policy which does not require any information about the valuations; however, such a policy is inventory-oblivious and performs poorly in computational experiments.

In \textbf{Section~\ref{sec::personalized}}, we discuss the final variant---\textit{personalized} pricing for \textit{stochastic} valuations---and show how our public pricing policies from Section~\ref{sec::stochastic} can be naturally adapted to exploit personal information, while maintaining the best-possible competitiveness of $\CR(\cP)$.

\subsection{Computational Experiments}

In \textbf{Section~\ref{sec::compExp}}, we conduct simulations in the personalized online revenue management setting popularized by \citet{GNR14}, where the \textit{stochastic} valuation distribution of each customer is revealed before a \textit{personalized} price is offered, but nothing is known about future customers.
For a fixed price set and starting inventory, we randomly generate 10000 sequences with different lengths and valuation distributions, that capture the full range of instances from ``myopic'' controls (which always maximize the immediate revenue) being optimal to ``conservative'' controls (which always offer the maximum price) being optimal.
We compare the average-case performance of our policy to those of \citet{BQ09,EM10}, as well as combinations and modifications of them designed to exploit personalized information, and the baseline ``myopic'' and ``conservative'' controls.
We repeat this experiment for different price ranges, price granularities (discrete vs.\ continuous prices), and staring inventory levels.
We draw the following conclusions:
\begin{enumerate}
\item Policies that make use of information about both
the valuation distributions and the remaining inventory
perform significantly better than those that do not, with awareness of remaining inventory being particularly important;
\item Among the two policies that do so,
(the personalized form of) our policy
accounts for inventory more precisely than
(the personalized form of) 
the booking limit policy of \citet{BQ09}, allowing it to achieve not only $\CR(\cP)$-competitiveness in theory, but improved average-case performance in practice.
\end{enumerate}
We should note that although we only tested policies designed for worst-case performance on an unknown demand sequence,
it is certainly plausible for a policy without such a worst-case guarantee to have good average-case performance.
However, we know of no such candidates, and have tried to exhaustively test all variants of known algorithms which operate under our framework (where it is not possible to forecast/learn future demand), and have found our Valuation Tracking procedure to perform best.
Developing alternative policies in this setting, not based on worst-case analysis, would be an interesting direction for future work.

\section{Literature Review}

Single-leg revenue management is a cornerstone problem in revenue management and pricing, as outlined in \citet{TvR06}.
Many different approaches for modeling demand have been considered over the years, as surveyed in \citet{AC11,db15}.
When nothing is known or can be learned about demand, one often resorts to competitive ratio analysis \citep{BEY05}, and this has been done for the single-item revenue management problem, where the decision is accept-reject, in \citet{BQ09}.
Our work extends their booking limit idea to pricing \citep[cf.][]{MM06}, by integrating it with the price-skimming idea from \citet{EM10} and its interpretation as a randomized price \citep{BS08}, in what we call a ``Valuation Tracking'' procedure.

We now compare our algorithmic ideas and results with other streams of literature.

\subsubsection*{Valuation tracking vs.\ prophet inequalities.}

The algorithmic idea of ``tracking'' what a clairvoyant would do and imitating it dates back to the ``prophet inequalities'' of \citet{krengel1977semiamarts}.
Prophet inequalities have seen a recent surge in the computer science literature \citep{kleinberg2012matroid,duetting2016revenue,duetting2017prophet,correa2017posted,correa2018pricing}, and like our paper, they also establish ratio guarantees relative to a clairvoyant optimum in pricing problems.

However, the crucial difference is that in the prophet inequalities literature, it is assumed that the universe of possible demand functions is given at the start (or can be sampled from, like in \citet{azar2014prophet}).
Even if the buyers can arrive in an ``adversarial'' order (chosen adaptively by the ``almighty'' adversary from \citet{kleinberg2012matroid}), it is guaranteed that each demand function given at the start will appear exactly once.
By contrast, our valuation tracking technique operates with only knowledge of the universe of \textit{prices}; it is not promised that there exists a customer who is willing to pay the highest price.
Our guarantees are also different in that they are parametrized by the given universe of prices $\cP$, while the classical prophet inequality result is a uniform $1/2$-guarantee.
Although our guarantee of $\CR(\cP)$ is usually smaller than $1/2$, it holds under less information about demand.

\subsubsection*{Valuation tracking vs.\ water-filling/water-level algorithms.}

After each customer, our valuation tracking procedure geometrically ``stacks'' her valuation above the $k$'th largest valuation seen thus far, in effect ``balancing'' the heights assigned to the $k$ units of inventory (see Section~\ref{sec::stacking}, Figure~\ref{fig::valuationConfig}).
This idea is similar to many algorithms which use ``water-filling'' to balance the levels of different buckets.
For example, in the Adwords problem, the buckets correspond to different advertisers' budgets, and the algorithm prioritizes assigning ads to buckets which are the lowest percent to being full \citep{KP00,MSVV07,DJK13}.
In e-commerce, a similar algorithm for personalized recommendation is used to make different items' inventories sell out at roughly the same time \citep{GNR14,CMSLX16,MSL17,cheung2018inventory}.
Online water-filling algorithms have also been used for allocation in Gaussian channels \citep{thekumparampil2014sub}.

Although the idea is similar, our valuation tracking procedure balances the valuations assigned to the \textit{different units of a single item}, while the above water-filling algorithms balance the \textit{allocation rates of different items}.
The constant competitive ratio guarantees in the above papers generally do not apply to our setting, because they don't consider the pricing decision for an item---the only exception is \citet{MSL17}.
However, our valuation tracking is a much more specialized procedure for a single item, and our competitive ratio guarantee $\CR(\cP)$ is strictly greater than their corresponding guarantee, for any price set $\cP$.
Our valuation tracking procedure also leads to public pricing algorithms, whereas they always assume the ability to personalize.

\subsubsection*{Valuation tracking vs.\ dynamic learning and pricing.}

Dynamic pricing (with inventory constraints) has been extensively studied in the setting where demand is initially unknown and must be learned over time.
The learning problem with stationary demand was initiated by \citet{besbes2009dynamic}, and has been subsequently studied by \citet{babaioff2011posting,badanidiyuru2012learning,badanidiyuru2013bandits,babaioff2015dynamic,den2015dynamic}, to list a few references.
Inventory-constrained dynamic learning and pricing has also been studied in various Bayesian \citep{araman2009dynamic,CW16}
and minimax \citep{lim2007relative,ZSQH16} settings.

Although these papers all consider posted-pricing decisions under limited inventory (called ``budgets'' in some cases), their main algorithmic challenge is in when to \textit{explore} new prices to learn unknown demand rates, vs.\ when to \textit{exploit} prices already known to generate a high revenue rate.
By contrast, for our valuation tracking algorithm, the adversarial demand cannot be learned, and the main algorithmic challenge in how to ``hedge'' against the different possibilities for the valuation that will arrive next.
The theoretical guarantees are also different---in learning and pricing, the focus is usually on how the ``regret'' (the \textit{difference} from the optimum) scales with the number of IID customers $T$; while in competitive analysis, the focus is on \textit{ratio} guarantees which hold without any assumptions on the number of customers $T$ or their valuations being IID.

\section{Algorithms for Public Pricing and Deterministic Valuations}\label{sec::deterministic}
In this section we consider the variant of the model introduced in Section~\ref{sec::setupIntro} with deterministic valuations and public pricing.  We assume a discrete set of prices but all of our results carry over to having a continuous range of prices taking the form $[\rmin,\rmax]$, as we discuss in Section~\ref{sec::continuum}.

For any positive integer $n$, let $[n]$ denote the set $\{1,\ldots,n\}$.
We assume that $\cP$ consists of $m$ discrete prices, i.e.\ $\cP=\{r^{(j)}:j\in[m]\}$, sorted $0<r^{(1)}<\ldots<r^{(m)}$.
We define $r^{(0)}$ to be 0, and then the valuation $V_t$ at any time $t$ lies in $\{r^{(0)},\ldots,r^{(m)}\}$, with $V_t=r^{(0)}$ representing the lack of a customer during time $t$.
Similarly, we define $r^{(m+1)}$ to be $\infty$, and then the price $P_t$ at any time $t$ lies in $\{r^{(1)},\ldots,r^{(m+1)}\}$, with $P_t=r^{(\infty)}$ representing the firm shutting off demand during time $t$, which is the only option if its inventory is out of stock.
Let $X_t$ be the indicator variable for making a sale during time $t$, i.e.\ it is 1 if $V_t\ge P_t$, and 0 otherwise.

Let $T$ denote the number of time steps.  None of the algorithms in this paper assume any knowledge of $T$; note that $T$ can always be made arbitrarily large by inserting customers with valuation 0.  We will hereafter treat $T$ as the unknown total number of customers, and interchangeably use the phrases ``customer $t$'' or ``time $t$'' to refer to valuation $V_t$ (even if it is 0).

In the public pricing variant, an \emph{online algorithm} must choose each $P_t$ based on only the history of past prices and valuations, $P_1,V_1,\ldots,P_{t-1},V_{t-1}$.  This history also determines the values of $X_1,\ldots,X_{t-1}$.  The online algorithm does not know $T$, and has no information about the current valuation $V_t$ or the future valuations $V_{t+1},\ldots,V_T$, when choosing $P_t$.  By contrast, the \emph{offline optimum} knows the entire sequence $V_1,\ldots,V_T$ before having to choose any prices.
Given any valuation sequence $V_1,\ldots,V_T$, we use the $P_t$ and $X_t$ variables to refer to the execution of an online algorithm on the valuation sequence.  Since the online algorithm may be randomized, we treat $P_t$ and $X_t$ as random variables.  Let $\ALG(V_1,\ldots,V_T)$ denote the total revenue earned by the online algorithm, equal to $\sum_{t=1}^TP_tX_t$.  Then $\bE[\ALG(V_1,\ldots,V_T)]$ is its expected revenue.  Meanwhile, let $\OPT(V_1,\ldots,V_T)$ denote the offline optimum for sequence $V_1,\ldots,V_T$, equal to the $\min\{k,T\}$ largest valuations from $V_1,\ldots,V_T$.
Formally, an online algorithm is said to be $c$-competitive if
\begin{align}
\bE[\ALG(V_1,\ldots,V_T)]\ge c\cdot\OPT(V_1,\ldots,V_T), && \forall\ T\ge1,(V_1,\ldots,V_T)\in(\cP\cup\{0\})^T. \label{eqn::compRatio}
\end{align}
For an initial setup given by $k$ and $\cP$, the maximum possible value of $c$ in~\eqref{eqn::compRatio} is called the \textit{competitive ratio}.
We will omit the arguments $(V_1,\ldots,V_T)$ in $\ALG$ and $\OPT$ when the context is clear.

As explained in Section~\ref{sec::setupIntro}, since our problem captures the problems of both \citet{BQ09} and \citet{EM10}, an upper bound for the value of $c$ in (\ref{eqn::compRatio}) is given by $\CR(\cP)$, as defined below.

\begin{definition}\label{def::q}
For any $m\ge1$, $0<r^{(1)}<\ldots<r^{(m)}$, and $\cP=\{r^{(1)},\ldots,r^{(m)}\}$, define:
\begin{itemize}
\item $q^{(j)}=1-\frac{r^{(j-1)}}{r^{(j)}}$ for all $j\in[m]$ (recall that $r^{(0)}=0$);
\item $q=\sum_{j=1}^mq^{(j)}$;
\item $\CR(\cP)=\frac{1}{q}$.
\end{itemize}
\end{definition}

The interpretation of $\frac{q^{(j)}}{q}$ in \citet{BQ09} is the fraction of initial inventory ``set aside'' for prices $j$ and higher.
The interpretation of $\frac{q^{(j)}}{q}$ in \citet{EM10} is the fraction of time that price $j$ should be charged.
Both of these papers establish that the competitiveness guarantee $c$ in~\eqref{eqn::compRatio} cannot be greater than $\CR(\cP)$, via Yao's minimax principle \citep{Yao77}.
In this paper, we derive $\CR(\cP)$-competitive algorithms for our dynamic pricing problem, which shows that the competitive ratio in fact equals $\CR(\cP)$.
%Therefore, for any fixed (but possibly randomized) online algorithm in our problem, there exists a sequence $V_1,\ldots,V_T$ such $\bE[\ALG(V_1,\ldots,V_T)]\le\CR(\cP)\cdot\OPT(V_1,\ldots,V_T)$.

\subsection{New Techniques, and why Existing Techniques Fail} \label{sec::attempts}
We explain the need for our main technical ingredient---a new ``valuation tracking'' procedure which incorporates both booking limits and price-skimming---by explaining why naive attempts to derive a $\CR(\cP)$-competitive algorithm fail under our dynamic pricing model.

We consider the following example, where the price set is $\cP=\{1,2,4\}$, and we will refer to customers with these valuations as being of type-L (Low), type-M (Medium), and type-H (High), respectively.
The competitive ratio for this price set derived by \citet{BQ09} and \citet{EM10} is $\CR(\cP)=1/2$.
We describe below their policies for this example.
\begin{itemize}
\item \underline{Booking Limits \citep{BQ09}}: Initially charge \$1; increase the price to \$2 after 1/2 of the starting inventory has been sold; further increase the price to \$4 after 3/4 of the starting inventory has been sold.  (This is the variant of booking limits with ``theft nesting''.)
\item \underline{Price-skimming \citep{EM10}}: Charge \$1 for 1/2 of the time steps; charge \$2 for 1/4 of the time steps; charge \$4 for 1/4 of the time steps.
\end{itemize}
These policies have the benefit that they never need any information about the valuations.
However, we now see why these policies fail to be $1/2$-competitive, under our dynamic pricing model.

\subsubsection*{Attempt 1: Direct implementation of booking limits.}
It is easy to see that this would not be 1/2-competitive---suppose just one type-H customer arrived at the start.  The algorithm would be charging the low price of \$1, while the offline optimum would be the customer's valuation of \$4.

Any direct implementation of price-skimming would suffer similarly, since there could be a single type-H customer who arrives during a time when the price is set to \$1.

\subsubsection*{Attempt 2: Price-skimming as a randomized price.}
It appears that the problem with Attempt~1 can be solved using the ``random price'' interpretation of price-skimming---instead of deterministically partitioning the time horizon according to ratios $\frac{1}{2},\frac{1}{4},\frac{1}{4}$ and offering prices 1,2,4 respectively, one could at each time step choose the prices randomly with respective probabilities $\frac{1}{2},\frac{1}{4},\frac{1}{4}$.  Then, if a single type-H customer arrives, the expected revenue would be
\begin{align*}
\frac{1}{2}\cdot1+\frac{1}{4}\cdot2+\frac{1}{4}\cdot4=2
\end{align*}
which is 1/2 of the customer's valuation of \$4.
It can be checked that 1/2 of the customer's valuation is also earned when it is \$1 or \$2; this is by construction of the price-skimming distribution.

However, having a fixed price-skimming distribution is no longer effective under inventory constraints.
Indeed, if a long sequence of type-L customers arrive, then this would deplete the inventory with high probability, and type-H customers who arrive last-minute would not be served, and the ratio of optimum earned would again be 1/4.

\subsubsection*{Attempt 3: Naive incorporation of booking limits into price-skimming.}
It appears that the problem with Attempt~2 can be solved by respecting the booking limits, i.e.\ forbidding price-skimming from randomly choosing the price of \$1 after 1/2 of the starting inventory has been sold.  However, this still fails to be 1/2-competitive, as shown by the following example.  Suppose the starting inventory is $k=4$, and that 2 type-H customers arrive followed by a type-L customer, with no customers arriving after that.  The optimum would be \$9.  However, the algorithm's revenue would only equal \$4: it would earn \$2 in expectation from each of the type-H customers, depleting 2 units of inventory, and then earn \$0 from the type-L customer due to the booking limit.

\subsubsection*{Our procedure: Valuation tracking.}
The problem with Attempt~3 leads to the following observation---the optimum is guaranteed to increase from the first 4 customers (since there are 4 units of inventory), so in order to be 1/2-competitive, the algorithm must maintain the initial price-skimming distribution for the first 4 customers.
After that, the algorithm can respect booking limits as long as customers rejected in this way would not increase the optimum, and in fact should do so, to avoid the problem in Attempt~2 of stocking out.
This motivates our procedure below.
\begin{itemize}
\item \underline{Valuation Tracking}:
At each time $t$, let $\ell_t$ denote the smallest value (possibly 0) in the 4 largest valuations to have arrived before time $t$.
%(where 4 is the starting inventory).
Then, randomly choose the price so that if the unknown valuation at time $t$ satisfies $V_t>\ell_t$, then the algorithm earns in expectation
\begin{align}
\frac{1}{2}(V_t-\ell_t). \label{eqn::introGuarantee}
\end{align}
\end{itemize}
In~\eqref{eqn::introGuarantee}, $V_t-\ell_t$ is the gain in the offline optimum should the valuation of customer $t$ be $V_t$, and 1/2 is the desired competitive ratio.
The constraint that the algorithm's revenue exactly equals (\ref{eqn::introGuarantee}) forces the algorithm to use the most \textit{inventory-conservative} control which maintains 1/2-competitiveness, thereby hedging against a stockout.
The price distribution used at each time $t$ depends on the inventory state, and in fact, the calculation for the algorithm's expected revenue must account for the probability of stocking out before time $t$.
The surprising fact is that it is \textit{possible} to choose price distributions (for each inventory state) which collectively guarantee expected revenue of (\ref{eqn::introGuarantee}).

\subsubsection*{Remark about personalization.}
In this paper we also consider the setting where a customer's valuation is distributionally-given \textit{before} her price is decided.
It is possible to modify the above examples to show that the attempts similarly fail in this setting---see E-supplement~\ref{sec::newEg}.

\subsection{Illustration of Valuation Tracking Procedure}\label{sec::stacking}

As explained above, the goal of Valuation Tracking is to earn a constant $\CR(\cP)$-fraction of the gain in $\OPT$ from each customer arriving.
This requires tracking the current value of $\OPT$, i.e. the sum of the $k$ largest valuations to have arrived before the current time step.
We now illustrate how this is done and why it is possible to be $\CR(\cP)$-competitive.

We consider the same example with $\cP=\{1,2,4\}$ and $\CR(\cP)=\frac{1}{2}$.  We consider a starting inventory of $k=5$, and suppose that 5 customers, with valuations $4,1,4,1,2$, have already arrived.  The current value of $\OPT$ is then the sum of these 5 valuations, $4+1+4+1+2=12$.

The procedure considers the possibilities for the increase in $\OPT$ from the next customer, which we denote as $\Delta\OPT$ (recall that a pricing decision must be made before knowing the valuation of the next customer).  Since the smallest valuation currently counted toward $\OPT$ is 1, if the valuation of the next customer if 4, then $\Delta\OPT=3$; if it is 2, then $\Delta\OPT=1$.  If the next customer has valuation not exceeding 1, then $\Delta\OPT=0$.  The procedure wants to guarantee that its \textit{expected} revenue on the next customer is at least $\frac{1}{2}\cdot\Delta\OPT$, for all of these possible valuations.  To accomplish this, it has to consider the probability that it has stocked out at this point; on those sample paths its revenue is 0.

Our procedure cleanly accounts for the probability of stocking out using the following approach.  Each customer is assigned to a \textit{specific} unit of inventory $i\in[k]$ upon arrival.  Each inventory unit $i$ maintains a variable $\level[i]$, which is the maximum valuation of a customer previously assigned to it.  The next customer is always assigned to an unit $i^*$ with the smallest value of $\level[i^*]$, \textit{regardless of whether that unit $i^*$ has already been sold}.  In this way, the assignment procedure is deterministic, and allows us to maintain an invariant: the probability a unit $i$ has been sold is dependent on only the (deterministic) value of $\level[i]$.

For each customer, the procedure makes an offer to her \textit{only if unit $i^*$ has not been sold}, at a random price exceeding $\level[i^*]$.  The higher $\level[i^*]$ is, the more likely it is that unit $i^*$ has been sold, and the lower the expected revenue from that customer.  However, if $\level[i^*]$ is high, then the potential increase in $\OPT$ from that customer is also lower; if the valuation of the customer does not exceed $\level[i^*]$, then both the procedure's revenue and $\Delta\OPT$ are 0.  By properly choosing the distributions for the random prices, our procedure is able to maintain the invariant on the probability of each unit being sold, while earning $\frac{1}{2}\cdot\Delta\OPT$ in expectation from each customer.

\begin{figure}
\begin{center}\begin{tikzpicture}
\draw(0,1)--(-\legendLength,1)node[left]{\footnotesize 1};
\draw(0,2)--(-\legendLength,2)node[left]{\footnotesize 2};
\draw(0,4)--(-\legendLength,4)node[left]{\footnotesize 4};
\draw(0.5,0)node[below]{\footnotesize 1};
\draw(1.5,0)node[below]{\footnotesize 2};
\draw(2.5,0)node[below]{\footnotesize 3};
\draw(3.5,0)node[below]{\footnotesize 4};
\draw(4.5,0)node[below]{\footnotesize 5};
\draw[->](0,0)--(0,4.33)node[above]{\footnotesize $\level[i]$};
\filldraw[opacity=1.0](0,0)rectangle(1,4);
\filldraw[opacity=.33](1,0)rectangle(2,1);
\filldraw[opacity=1.0](2,0)rectangle(3,4);
\filldraw[opacity=.33](3,0)rectangle(4,1);
\filldraw[opacity=.67](4,0)rectangle(5,2);
\node at (2.5,5){\large Before};
\node at (2.5,-.75){\footnotesize inventory unit $i$};
%\node[align=center] at (-1,-.75){\footnotesize Inventory Unit\\[-16.5pt]\footnotesize Available?};
%\node at (0.5,-.75){\xmark};
%\node at (1.5,-.75){\cmark};
%\node at (2.5,-.75){\xmark};
%\node at (3.5,-.75){\xmark};
%\node at (4.5,-.75){\cmark};

\draw(\beforeAfterShift,1)--(\beforeAfterShift-\legendLength,1)node[left]{\footnotesize 1};
\draw(\beforeAfterShift,2)--(\beforeAfterShift-\legendLength,2)node[left]{\footnotesize 2};
\draw(\beforeAfterShift,4)--(\beforeAfterShift-\legendLength,4)node[left]{\footnotesize 4};
\draw(\beforeAfterShift+0.5,0)node[below]{\footnotesize 1};
\draw(\beforeAfterShift+1.5,0)node[below]{\footnotesize 2};
\draw(\beforeAfterShift+2.5,0)node[below]{\footnotesize 3};
\draw(\beforeAfterShift+3.5,0)node[below]{\footnotesize 4};
\draw(\beforeAfterShift+4.5,0)node[below]{\footnotesize 5};
\draw[->](\beforeAfterShift,0)--(\beforeAfterShift,4.33)node[above]{\footnotesize $\level[i]$};
\filldraw[opacity=1.0](\beforeAfterShift,0)rectangle(\beforeAfterShift+1,4);
\filldraw[opacity=.33](\beforeAfterShift+1,0)rectangle(\beforeAfterShift+2,1);
\filldraw[opacity=.67](\beforeAfterShift+1,1)rectangle(\beforeAfterShift+2,2);
\filldraw[opacity=1.0](\beforeAfterShift+2,0)rectangle(\beforeAfterShift+3,4);
\filldraw[opacity=.33](\beforeAfterShift+3,0)rectangle(\beforeAfterShift+4,1);
\filldraw[opacity=.67](\beforeAfterShift+4,0)rectangle(\beforeAfterShift+5,2);
\node at (\beforeAfterShift+2.5,5){\large After};
\node at (\beforeAfterShift+2.5,-.75){\footnotesize inventory unit $i$};
%\node[align=center] at (\beforeAfterShift-1,-.75){\footnotesize Inventory Unit\\[-16.5pt]\footnotesize Available?};
%\node at (\beforeAfterShift+0.5,-.75){\xmark};
%\node at (\beforeAfterShift+1.5,-.75){\cmark};
%\node at (\beforeAfterShift+2.5,-.75){\xmark};
%\node at (\beforeAfterShift+3.5,-.75){\xmark};
%\node at (\beforeAfterShift+4.5,-.75){\cmark};

\draw(5+\beforeAfterShift/2-3,2)--(5+\beforeAfterShift/2-3-\legendLength,2)node[left]{\footnotesize 2};
\filldraw[opacity=.67](5+\beforeAfterShift/2-3,0)rectangle(5+\beforeAfterShift/2-2,2);
\draw[->,ultra thick](5+\beforeAfterShift/2-2.5,1.5)--(1.5,1.5);
\node[align=center] at (5+\beforeAfterShift/2-2.5,2.75){Revealed\\[-11pt]Valuation};
\end{tikzpicture}\end{center}
\caption{The configuration of valuations, before and after a customer with valuation~2 arrives.}
\label{fig::valuationConfig}
\end{figure}
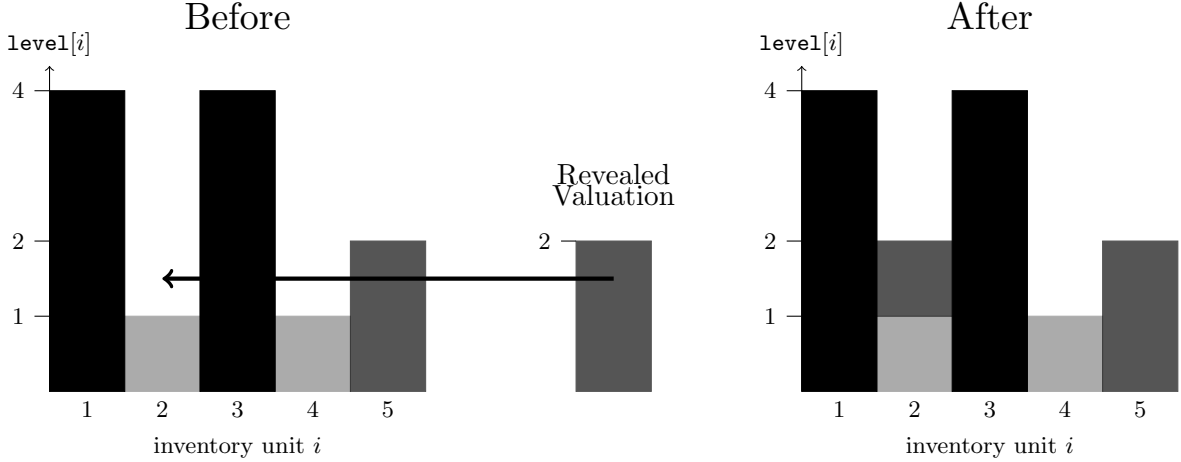

Returning to the example, given that the first 5 customers had valuations $4,1,4,1,2$, the values of $\level[i]$ for $i=1,\ldots,k$ are shown in the LHS of Fig.~\ref{fig::valuationConfig}.  The next customer, ``customer~\#6'', is assigned to inventory unit~2.  After her valuation is revealed to be 2, the updated configuration is shown on the RHS of Fig.~\ref{fig::valuationConfig}, regardless of whether she was offered a price.

Customer~\#6 would have been turned away if unit~2 was sold before her arrival, even if other units were available.  When $\cP=\{1,2,4\}$, the probability that a unit $i$ has been sold equals $0,\frac{1}{2},\frac{3}{4},1$ if $\level[i]$ is $0,1,2,4$, respectively.  These probabilities correspond to the values of $q^{(j)}$ from Definition~\ref{def::q}.  Since $\level[2]$ was 1 before customer~\#6 arrived, she is made an offer with probability $\frac{1}{2}$, at a random price exceeding 1.  The price is 2 with probability proportional to $\frac{1}{4}$, and 4 with probability proportional to $\frac{1}{4}$ (again using the values of $q^{(j)}$), hence each price would be offered with probability $\frac{1}{2}$.  The customer's valuation is 2, so she will only buy the item if offered price~2, which occurs with total probability $\frac{1}{2}\cdot\frac{1}{2}=\frac{1}{4}$.  Note that:
\begin{enumerate}
\item Customer~\#6 increases the probability of unit~2 being sold from $\frac{1}{2}$ to $\frac{3}{4}$, which is consistent with her increasing $\level[2]$ from 1 to 2;
\item Customer~\#6 increases the value of $\OPT$, equal to $\sum_{i=1}^k\level[i]$, by 1 (from 12 to 13);
\item Customer~\#6 brings in expected revenue $\frac{1}{4}\cdot2=\frac{1}{2}$.
\end{enumerate}
Therefore, during time step~6, our procedure has earned expected revenue $\frac{1}{2}\cdot\Delta\OPT$.  We will show that it achieves this for a general $\cP$, and all time steps $t$, regardless of the valuation of customer $t$.

\subsection{Valuation Tracking Procedure and Analysis}\label{sec::valuationTrackingProc}
We now formalize our valuation tracking procedure, in Algorithm~\ref{alg::valuationTrackingProc}.

\SingleSpacedXI
\begin{algorithm2e}[t]
\SetAlgoNoLine
\KwIn{Customers $t=1,2,\ldots$ arriving online, with each valuation $V_t$ revealed after the price $P_t$ is chosen.}
\KwOut{For each customer $t$, a (possibly random) price $P_t$ for her.}
$\level[i]=0,\sold[i]=\false$ for $i=1,\ldots,k$\;
$t=1$\;
\While{customer $t$ arrives}{
$i^*_t=\argmin_{i}\level[i]$\;
set $\ell_t$ to the index in $\{0,\ldots,m\}$ such that $\level[i^*_t]=r^{(\ell_t)}$\;
\eIf{$\sold[i^*_t]=\false$}{\label{line::if_start}
let $\qtot=\sum_{j=\ell_t+1}^mq^{(j)}$\;
randomly choose a price from $\{r^{(\ell_t+1)},\ldots,r^{(m)}\}$, with probabilities $\frac{q^{(\ell_t+1)}}{\qtot},\ldots,\frac{\ell_t+1}{\qtot}$ respectively (note that these probabilities sum to 1)\;\label{line::sum_to_one}
}{
reject the customer by choosing price $\infty$\;
}\label{line::if_end}
observe valuation $V_t$ and purchase decision $X_t$\;
$\level[i^*_t]=\max\{\level[i^*_t],V_t\}$\;\label{line::increment_anyway}
\If{$X_t=1$}{
$\sold[i^*_t]=\true$\;
}
$t=t+1$\;
}
\caption{Valuation Tracking Procedure}
\label{alg::valuationTrackingProc}
\end{algorithm2e}
\DoubleSpacedXI

In line~\ref{line::sum_to_one}, the procedure offers exactly one of the prices $r^{(\ell_t+1)},\ldots,r^{(m)}$, with the offering probabilities summing to unity.  Note that it cannot branch to line~\ref{line::sum_to_one} if $\ell_t=m$.  This can be seen in the following way.  If $\ell_t=m$, then $i^*_t$ must have been assigned to some past customer $t'$ with $V_{t'}=r^{(m)}$.  At time $t'$, either inventory unit $i^*_t$ was already sold, or customer $t'$ was offered a price at most $r^{(m)}$, which she would have accepted.  In either case, $\sold[i^*_t]$ must be $\true$.

The analysis of Algorithm~\ref{alg::valuationTrackingProc} is conceptually simple.
Let $b_{i,t}$ be the index in $0,\ldots,m$ such that $\level[i]=r^{(b_{i,t})}$ at the end of time $t$, and $j_t$ be the index such that $V_t=r^{(j_t)}$.
We show that the following claims are maintained:
\begin{enumerate}
\item At the end of each time step $t$, the probability that any inventory unit $i$ has been sold is $(\sum_{j=1}^{b_{i,t}}q^{(j)})/q$;
\item During a time step $t$, if the valuation of the customer exceeds the level of the inventory unit she is assigned to, i.e. $j_t>\ell_t$, then:
\begin{enumerate}
\item The expected revenue earned by Algorithm~\ref{alg::valuationTrackingProc} is $\frac{1}{q}(r^{(j_t)}-r^{(\ell_t)})$;
\item The increase in the offline optimum is $r^{(j_t)}-r^{(\ell_t)}$.
\end{enumerate}
\end{enumerate}
If $j_t\le\ell_t$ during a time step $t$, then both the revenue of Algorithm~\ref{alg::valuationTrackingProc} and the gain in $\OPT$ are 0.

These claims establish the following theorem, whose full proof is deferred to E-supplement~\ref{sec::valuationTrackingProcPf}.

\begin{theorem}\label{thm::valuationTrackingProc}
Algorithm~\ref{alg::valuationTrackingProc} is $\CR(\cP)$-competitive.
\end{theorem}

\subsection{Modified Algorithm based on Valuation Tracking Procedure}\label{sec::modifiedAlg}
In this section we present a modified version of Algorithm~\ref{alg::valuationTrackingProc} which is useful for the subsequent developments under the stochastic-valuation model in Section~\ref{sec::stochastic}.

First we show how to modify Algorithm~\ref{alg::valuationTrackingProc} so that its decision at each time $t$ depends on only the remaining inventory, instead of the entire history of purchase decisions $X_1,\ldots,X_{t-1}$.
\begin{definition}
For all $t=0,\ldots,T$, let $I_t$ denote the random variable for the amount of remaining inventory at the end of time $t$, which is equal to $k-\sum_{t'=1}^tX_{t'}$.
\end{definition}

Our modified algorithm makes an offer to customer $t$ according to the \textit{probability} that unit $i^*_t$ hasn't been sold, conditioned on the realized value of $I_{t-1}$.  In this way, its decisions depend on only the inventory state, instead of the exact decisions of past customers.
\begin{definition}[Algorithm~\ref{alg::valuationTrackingProc}']
Define the following algorithm for choosing the price at each time $t$, based on the past valuations $V_1,\ldots,V_{t-1}$ and the amount of remaining inventory $I_{t-1}$.
\begin{enumerate}
\item Consider the indices $i^*_t$ and $\ell_t$ during iteration $t$ of Algorithm~\ref{alg::valuationTrackingProc}, which are deterministic based on $V_1,\ldots,V_{t-1}$.
\item Compute the probability that $\sold[i^*_t]=\true$ on a run of Algorithm~\ref{alg::valuationTrackingProc}, \textit{conditioned on $I_{t-1}$ units of inventory remaining after time $t-1$} in that run.  Let $\gamma_t$ denote this probability.
\label{line::dpCompute}
\item With probability $\gamma_t$, make an offer to customer $t$ with the same price distribution as Algorithm~\ref{alg::valuationTrackingProc} (line~\ref{line::sum_to_one}); with probability $1-\gamma_t$, offer price $\infty$ to customer $t$.
\label{line::turnAway}
\end{enumerate}
\end{definition}

Algorithm~\ref{alg::valuationTrackingProc}' chooses the distribution for $P_t$ by ``averaging'' over all runs of Algorithm~\ref{alg::valuationTrackingProc} which have the same value of $I_{t-1}$.  We first remark that this can be done in polynomial time, despite there being exponentially many sample paths for Algorithm~\ref{alg::valuationTrackingProc}.  We prove the following in E-supplement~\ref{sec::modifiedAlgPf}.
\begin{lemma}\label{lem::polytime_dp}
The value of $\gamma_t$ in Step~\ref{line::dpCompute} of Algorithm~\ref{alg::valuationTrackingProc}' can be computed in polynomial time.
\end{lemma}

We now introduce some notation to disambiguate between random variables depicting the runs of different algorithms.
\begin{definition}\label{def::diff_algs}
For an algorithm $\cA$, let $P^{\cA}_t$, $X^{\cA}_t$, and $I^{\cA}_t$ be the random variables for the price at time $t$, purchase decision at time $t$, and inventory remaining at the end of time $t$, respectively.  Let $\ALG^{\cA}$ be the random variable for the total revenue earned by algorithm $\cA$.  We will omit the superscripts $\cA$ when the context is clear.

Let $\cA=\VK$ refer to Algorithm~\ref{alg::valuationTrackingProc} and $\cA=\VKp$ refer to Algorithm~\ref{alg::valuationTrackingProc}'.
\end{definition}

We show that Algorithms~\ref{alg::valuationTrackingProc} and \ref{alg::valuationTrackingProc}' are virtually the same in that they have identical distributions for the remaining inventory at each time step, as well as the random price at each time step conditioned on any value of remaining inventory.  This also establishes that Algorithm~\ref{alg::valuationTrackingProc}' is feasible, in that it does not try to make a sale with zero remaining inventory.
\begin{lemma}\label{lem::detVInd}
For all $t\in[T]$, $k'\in\{0,\ldots,k\}$ such that $\Pr[I^{\VKp}_{t-1}=k']>0$, and $j\in\{1,\ldots,m,m+1\}$, $\Pr[\pvkp_t=r^{(j)}|I^{\VKp}_{t-1}=k']=\Pr[\pvk_t=r^{(j)}|I^{\VK}_{t-1}=k']$.

Also, for all $t=0,\ldots,T$ and $k'\in\{0,\ldots,k\}$, $\Pr[I^{\VKp}_t=k']=\Pr[I^{\VK}_t=k']$.
\end{lemma}

Lemma~\ref{lem::detVInd}, proven in E-supplement~\ref{sec::modifiedAlgPf}, is a consequence of the design of Algorithm~\ref{alg::valuationTrackingProc}'.  For all $t$, the random price $\pvkp_t$ is identically distributed as $\pvk_t$, conditional on any value for the amount of remaining inventory at the end of time $t-1$.  Hence if $I^{\VKp}_{t-1}$ and $I^{\VK}_{t-1}$ are identically distributed, then so are $I^{\VKp}_t$ and $I^{\VK}_t$.  This allows us to inductively establish that the two algorithms have the same aggregate behavior after combining all sample paths, even though their behavior may differ given a specific history of purchase decisions.
This also makes it easy to see that the expected revenues of the two algorithms are the same.  Lemma~\ref{lem::detVInd} directly implies the following theorem.
\begin{theorem}\label{thm::modifiedAlg}
Algorithm~\ref{alg::valuationTrackingProc}' is $\CR(\cP)$-competitive.
\end{theorem}

\subsection{Further Modified Algorithm and Structural Properties}\label{sec::structural}
In this section we present a further-modified version of Algorithm~\ref{alg::valuationTrackingProc}' which satisfies two structural properties: (i) it never offers price $\infty$ to a customer if it has remaining inventory, offering the maximum price instead; (ii) the distribution of prices offered to a customer is \textit{strictly stochastically-decreasing} (see Corollary~\ref{cor::stoc_decr}) in the amount of remaining inventory.
%Property~(ii) is relevant for the personalized revenue management model in Section~\ref{sec::personalized}, because it defines the tradeoff between inventory and revenue at each time step.
Property~(ii) is the stochastic analogue of the classical structural result from \citet[Thm.~1]{GvR94} and its generalization to non-homogeneous demand in \citet[Thm.~3]{ZZ00}: at any time step, if the firm has more inventory, then the optimal offering price is strictly lower.
\begin{definition}[Algorithm~\ref{alg::valuationTrackingProc}'']
Define the following modification to Algorithm~\ref{alg::valuationTrackingProc}': in Step~\ref{line::turnAway}, offer price the maximum price $r^{(m)}$ to customer $t$, instead of offering $\infty$, with probability $1-\gamma_t$.
\end{definition}

We prove the following general lemma, which is intuitively easy to see, in E-supplement~\ref{sec::modifiedAlgPf}.
\begin{lemma}\label{lem::strictDominance}
Let $\cA$ be any pricing algorithm.  Let $\cA'$ be the modified algorithm which: whenever $\cA$ would offer price $\infty$ to a customer while there is remaining inventory, $\cA'$ offers price $r^{(m)}$ instead.  Then on any valuation sequence $V_1,\ldots,V_T$, $\bE[\ALG^{\cA'}]\ge\bE[\ALG^{\cA}]$.
\end{lemma}

Lemma~\ref{lem::strictDominance} shows that Algorithm~\ref{alg::valuationTrackingProc}'' is $\CR(\cP)$-competitive.  Now, we would like to further show that the probability of Algorithm~\ref{alg::valuationTrackingProc}' rejecting, or correspondingly the probability of Algorithm~\ref{alg::valuationTrackingProc}'' offering the maximum price, is smaller when conditioned on larger values of remaining inventory.
\begin{theorem}\label{thm::structural}
Suppose that the unconditional probability of Algorithm~\ref{alg::valuationTrackingProc}' rejecting customer $t$, $\Pr[\sold[i^*_t]=\true]$, lies in (0,1).
Then for any $k_1<k_2$ with $\Pr[I_{t-1}=k_1]>0$ and $\Pr[I_{t-1}=k_2]>0$,
\begin{align*}
\Pr[\sold[i^*_t]=\true|I_{t-1}=k_1]>\Pr[\sold[i^*_t]=\true|I_{t-1}=k_2].
\end{align*}
\end{theorem}

This structural property is intuitive, and we defer its proof to E-supplement~\ref{sec::modifiedAlgPf}.
%The price distributions chosen by Algorithm~\ref{alg::valuationTrackingProc} for each combination of time and remaining inventory will be useful in Section~\ref{sec::personalized}.
Theorem~\ref{thm::structural} allows us to conclude with the following corollary about strict stochastic dominance.
\begin{corollary}\label{cor::stoc_decr}
For any $t$, suppose $k_1<k_2$ with $\Pr[I_{t-1}=k_1]>0,\Pr[I_{t-1}=k_2]>0$, and let $F_{P_t|I_{t-1}}(x|k_1),F_{P_t|I_{t-1}}(x|k_2)$ denote the CDF's of the price $P_t$ offered by Algorithm~\ref{alg::valuationTrackingProc}'' at time $t$ conditional on the remaining inventory $I_{t-1}$ being $k_1,k_2$, respectively.  Then $F_{P_t|I_{t-1}}(x|k_1)\le F_{P_t|I_{t-1}}(x|k_2)$ for all $x\in\bR$, and moreover, the dominance is strict in that $F_{P_t|I_{t-1}}(x|k_1)<F_{P_t|I_{t-1}}(x|k_2)$ for some price $x\in\cP$.
\end{corollary}

This intuitive property of raising prices as inventory dwindles contributes to the strong empirical performance of our algorithm, as we will explain in Section~\ref{sec::compExp}.

\section{Algorithms for Public Pricing and Stochastic Valuations}\label{sec::stochastic}
In this section we extend the results from Section~\ref{sec::deterministic} to allow for stochastic valuations.
The model with stochastic valuations differs from the model with deterministic valuations studied in the previous section in the following ways.
\begin{itemize}
\item The valuation of each arriving customer is now \textit{randomly} drawn from some distribution.  The valuations of different customers are independent, but not necessarily identically distributed.
\item An online public pricing algorithm is given the valuation \textit{distribution} of each customer after the price for that customer has been chosen.
\item The algorithm also sees the purchase decision of the customer at the price offered, and hence can posterior-update this valuation distribution for the customer if it wishes.
\end{itemize}

\begin{definition}
We use the following notation, defined for all $t$:
\begin{itemize}
\item $V_t$: the valuation of customer $t$, a random variable taking values in $\{r^{(0)},r^{(1)},\ldots,r^{(m)}\}$;
\item $\vu_t$: the probability vector $(v^{(0)}_t,v^{(1)}_t,\ldots,v^{(m)}_t)$ for the distribution of $V_t$, with $v^{(j)}_t=\Pr[V_t=r^{(j)}]$ and $\sum_{j=0}^mv^{(j)}_t=1$.
%\item $\vV_t$: $(V_1,\ldots,V_t)$, the vector of realized valuations up to time $t$;
%\item $\vP^{\cA}_t$: $(P^{\cA}_1,\ldots,P^{\cA}_t)$, the vector of prices up to time $t$ chosen by algorithm $\cA$;
\end{itemize}
\end{definition}

With stochastic valuations, we say that an online algorithm is $c$-competitive if
\begin{align} \label{eqn::competitiveStoch}
\frac{\bE_{V_1\sim\vu_1,\ldots,V_T\sim\vu_T}[\bE[\ALG(V_1,\ldots,V_T)]]}{\bE_{V_1\sim\vu_1,\ldots,V_T\sim\vu_T}[\OPT(V_1,\ldots,V_T)]}\ge c, && \forall\ T\ge1,(\vu_1,\ldots,\vu_T).
\end{align}
In~\eqref{eqn::competitiveStoch},
the numerator is the revenue earned in expectation over both the randomness in the valuations and any randomness in the algorithm, while
the denominator is the clairvoyant optimum defined as the expected sum of the $k$ largest valuations.
We note that this optimum is an upper bound on the revenue of the optimal dynamic program knowing distributions $\vu_1,\ldots,\vu_T$ in advance, and hence any value of $c$ which satisfies~\eqref{eqn::competitiveStoch} implies the same guarantee relative to the optimal DP.
Moreover, since the guarantee $c$ in~\eqref{eqn::competitiveStoch} cannot be greater than $\CR(\cP)$ even when $V_1,\ldots,V_T$ are deterministic (in which case the DP value equals $\OPT(V_1,\ldots,V_T)$), our $\CR(\cP)$-competitive algorithms are still optimal if one were to use the DP value in the denominator of definition~\eqref{eqn::competitiveStoch} instead.
By using the relaxation $\bE_{V_1\sim\vu_1,\ldots,V_T\sim\vu_T}[\OPT(V_1,\ldots,V_T)]$, our denominator is much easier to analyze.
%That is, although we have relaxed the optimum and allowed it to know the valuations in advance for simplicity of analysis, we have not diminished the competitive ratio through definition~\ref{eqn::competitiveStoch}.

We should point out that with stochastic valuations, another commonly-studied relaxation which could have been used in the denominator of~\eqref{eqn::competitiveStoch} is the deterministic linear program (DLP) with distributions $\vu_1,\ldots,\vu_T$.  By the same argument about $V_1,\ldots,V_T$ being deterministic, the competitive ratio relative to the DLP also cannot be greater than $\CR(\cP)$.  However, we provide in E-supplement~\ref{sec::dlp} examples of $\cP$ where the competitive ratio relative to the DLP is strictly smaller than $\CR(\cP)$.
Therefore, we choose our relaxation because it automatically provides the tight competitive ratio relative to the DP.
Nonetheless, we believe that pinpointing the different value of competitive ratio relative to the DLP is an interesting theoretical problem.

\subsection{Optimally-Competitive Algorithm with Exponential Runtime}\label{sec::exp_time}
Having defined our clairvoyant optimum, we now derive $\CR(\cP)$-competitive algorithms in the stochastic-valuation model.
We do so by using our valuation tracking procedure as a subroutine, in a similar way to the development in Section~\ref{sec::modifiedAlg}, which may be helpful to reference.

Conceptually, our algorithm is a generalization of Algorithm~\ref{alg::valuationTrackingProc}' to stochastic valuations.  However, since the assignment procedure in Algorithm~\ref{alg::valuationTrackingProc} is no longer deterministic, we describe the algorithm the following way.  At a time step $t$, having seem distributions $\vu_1,\ldots,\vu_{t-1}$ and knowing remaining inventory $k'$:
\begin{enumerate}
\item Consider a run of Algorithm~\ref{alg::valuationTrackingProc} to the end of time $t$, where $V_1,\ldots,V_{t-1}$ are randomly drawn according to $\vu_1,\ldots,\vu_{t-1}$.  For all $j\in\{1,\ldots,m,m+1\}$, compute $\Pr[\pvk_t=r^{(j)}|I^{\VK}_{t-1}=k']$, where the probability is over both the random valuations, and the random prices chosen by the algorithm. (If $\Pr[I^{\VK}_{t-1}=k']$ has measure 0, then choose price $r^{(m+1)}$.) \label{step_again}
\item For each $j\in\{1,\ldots,m,m+1\}$, choose price $r^{(j)}$ with probability $\Pr[\pvk_t=r^{(j)}|I^{\VK}_{t-1}=k']$.
\end{enumerate}
We will call this algorithm $\Exp$ and use random variables $P^{\Exp}_t,I^{\Exp}_t$ (coming from Definition~\ref{def::diff_algs}) to refer to its execution.
It will be seen that $\Exp$ is a feasible policy when we establish that $I^{\Exp}_t$ and $I^{\VK}_t$ are identically distributed at every time step $t$.

We allow $\Exp$ to use Exponential running time, which is required for computing the conditional probabilities $\Pr[\pvk_t=r^{(j)}|I^{\VK}_{t-1}=k']$.
To see why, note that the geometric stacking configuration from Algorithm~\ref{alg::valuationTrackingProc} is no longer fixed once the valuations are stochastic.
Therefore, our polynomial-time computation trick (Lemma~\ref{lem::polytime_dp}) for Algorithm~\ref{alg::valuationTrackingProc}' (the special case of $\Exp$ under deterministic valuations) can no longer be used.  Instead, we compute $\Pr[\pvk_t=r^{(j)}|I^{\VK}_{t-1}=k']$ by brute force, enumerating all histories of prices offered $\pvk_1,\ldots,\pvk_{t-1}$ and valuations realized $V_1,\ldots,V_{t-1}$ which would lead to $I^{\VK}_{t-1}=k'$.
Each such history implies a stacking configuration based on $V_1,\ldots,V_{t-1}$, from which the random price distribution of Algorithm~\ref{alg::valuationTrackingProc} can be computed.
Taking a weighted average over these histories tells us the probability with which $P^{\VK}_t$ would equal each price $r^{(j)}$.

%Unfortunately, at each time step $t$, this takes time exponential in $t$.  The computational difficulty arises because the assignment procedure in  is no longer deterministic, as it was throughout Section~\ref{sec::deterministic}.

For now, we ignore computational considerations and show that the exponential-time algorithm $\Exp$ is $\CR(\cP)$-competitive as an online algorithm; in Section~\ref{sec::poly_time} we show how we can sample from the conditional distribution of $P^{\VK}_t$ given $I^{\VK}_{t-1}=k'$, to achieve polynomial runtime while only losing $\ve$ in the competitiveness.
The following lemma is analogous to Lemma~\ref{lem::detVInd} and proved in E-supplement~\ref{sec::stochasticPf}.
\begin{lemma}\label{lem::exp_states}
For all $t\in[T]$, $k'\in\{0,\ldots,k\}$ such that $\Pr[I^{\Exp}_{t-1}=k']>0$, and $j\in\{1,\ldots,m,m+1\}$,
\begin{equation}\label{eqn::pricing}
\Pr[\pexp_t=r^{(j)}|I^{\Exp}_{t-1}=k']=\Pr[\pvk_t=r^{(j)}|I^{\VK}_{t-1}=k'].
\end{equation}
Also, for all $t=0,\ldots,T$ and $k'\in\{0,\ldots,k\}$,
\begin{equation}\label{eqn::inventory}
\Pr[I^{\Exp}_t=k']=\Pr[I^{\VK}_t=k'].
\end{equation}
\end{lemma}

Lemma~\ref{lem::exp_states} establishes that $\Exp$ is a feasible policy, i.e.\ it does not try to make a sale with no inventory remaining.
%Indeed, we know that Algorithm~\ref{alg::valuationTrackingProc} is feasible, so $\Pr[\pvk_t<\infty|I^{\VK}_{t-1}=0]=0$.  By (\ref{eqn::pricing}), $\Pr[\pexp_t<\infty|I^{\Exp}_{t-1}=0]=0$ as well.
%Similarly, Lemma~\ref{lem::exp_states} implies that Step~\ref{step_again} of $\Exp$ will never be executed when $ $ has measure 0.
Having established this, it remains to prove that $\Exp$ is optimally competitive.  Theorem~\ref{thm::exp} is proved in E-supplement~\ref{sec::stochasticPf}.
\begin{theorem}\label{thm::exp}
$\bE[\ALG^{\Exp}(V_1,\ldots,V_T)]=\bE[\ALG^{\VK}(V_1,\ldots,V_T)]$.  By Theorem~\ref{thm::valuationTrackingProc}, $\ALG^{\VK}(V_1,\ldots,V_T)=\frac{1}{q}\OPT(V_1,\ldots,V_T)$ for all realizations $(V_1,\ldots,V_T)$.  Therefore, $\Exp$ is $\CR(\cP)$-competitive.
\end{theorem}

We should point out that although $\Exp$ does not inherit the polynomial-time property from Algorithm~\ref{alg::valuationTrackingProc}', it does inherit the structural property of the price at any time being stochastically-decreasing in the amount of remaining inventory.  This is immediate from Theorem~\ref{thm::structural}, which holds conditioned on any realization of $V_1,\ldots,V_T$.

Also, note the following.  $\pexp_t$ and $\pvk_t$ are only guaranteed to be identically distributed when averaged over \emph{all} the sample paths up to time $t-1$ such that the total remaining inventory is $k'$.  They may not be identically distributed when conditioned on a specific purchase sequence $X_1,\ldots,X_{t-1}$ such that $\sum_{t'=1}^{t-1}X_{t'}=k-k'$, or a specific valuation sequence $V_1,\ldots,V_{t-1}$.  Nonetheless, our method works in general.  For example, if valuations were correlated, then we would condition on both $I_{t-1}$ and $V_1,\ldots,V_{t-1}$.  One benefit of conditioning on only $I_{t-1}$ in the independent case is to limit the state space, which is necessary for our polynomial-time sampling algorithm in Section~\ref{sec::poly_time}.

\subsection{Emulating the Exponential-runtime Algorithm using Sampling}\label{sec::poly_time}
In this section we show how to ``emulate'' $\Exp$, using sampling, to achieve a polynomial runtime.  First we provide a high-level overview of the challenges and the techniques used to overcome them.
%The intrinsic difficulty is that our original procedure is based on tracking the value of the offline optimum, but this becomes a \#P-hard problem when the optimum equals the expected value of the $k$ largest elements from independent realizations (see \citet{Hag88}).
%this is computing the expected project duration in a PERT network with independent task durations---

First, suppose we are at the start of time $t$, with inventory $k'$ remaining.  If we randomly sample a run of Algorithm~\ref{alg::valuationTrackingProc} (drawing valuations randomly) such that $I^{\VK}_{t-1}=k'$, and copy price $P^{\VK}_t$ for time $t$, then we would match the probabilities prescribed $\Exp$.  This motivates the following algorithm: sample runs of Algorithm~\ref{alg::valuationTrackingProc} to the end of $t-1$ until hitting one where $I^{\VK}_{t-1}=k'$, and then choose the price for time $t$ according to lines~\ref{line::if_start}--\ref{line::if_end} of Algorithm~\ref{alg::valuationTrackingProc}.  Such an algorithm is equivalent to $\Exp$, and thus would be $\CR(\cP)$-competitive.

However, on sample paths where $\Pr[I^{\VK}_{t-1}=k']$ is small, the sampling could take arbitrarily long.  We limit the number of sampling tries so that the algorithm deterministically finishes in polynomial time, and show that the total measure of sample paths which fail at any point is $O(\ve)$.
Unfortunately, there could be correlation between the sampling failing, and having high revenue on a sample path.  Nonetheless, we can \emph{couple} the sample paths of the sampling algorithm to those of the exponential-time algorithm, mark the first point of failure on each sample path, and bound the difference in revenue after that point.

The details of the sampling algorithm, which we will call $\Samp$, are specified in Algorithm~\ref{alg::sampling}.
Note that in line~\ref{line::define_C} of the algorithm, $C$ is a positive integer to be chosen later.  The decision of what to do when the sampling \emph{fails}, i.e.\ defaults to line~\ref{line::default}, is inconsequential, since in our analysis we do not expect any revenue from a sample path \emph{after} the first point of failure.

\SingleSpacedXI
\begin{algorithm2e}[t]
\SetAlgoNoLine
\KwIn{Customers $t=1,2,\ldots$ arriving online, with each valuation distribution $\vu_t$ revealed after the price $P_t$ is chosen.}
\KwOut{For each customer $t$, a (possibly random) price $P_t$ for her.}
$\inventory=k$\;
$t=1$\;
\While{customer $t$ arrives}{
\Repeat{$C(k+1)t^2$ runs elapse}{
run Algorithm~\ref{alg::valuationTrackingProc} to the start of time $t$, with valuations $V_1,\ldots,V_{t-1}$ drawn according to $\vu_1,\ldots,\vu_{t-1}$, and prices $\pvk_1,\ldots,\pvk_{t-1}$ realized according to the random prices chosen by Algorithm~\ref{alg::valuationTrackingProc}\;
\If{$I^{\VK}_{t-1}=\inventory$}{
choose each price $r^{(1)},\ldots,r^{(m)},r^{(m+1)}$ according to the probability that Algorithm~\ref{alg::valuationTrackingProc} (on this run) would choose that price for customer $t$\;\label{line::hit_start}
observe $\vu_t$\;
observe purchase decision of customer $t$ and update $\inventory$ accordingly\;
$t=t+1$ and \textbf{continue} to next iteration of \textbf{while} loop\;\label{line::hit_end}
}
}\label{line::define_C}
choose price $\infty$\;\label{line::default}
$t=t+1$\;
}
\caption{Stochastic Sampling Algorithm based on Valuation Tracking Procedure}
\label{alg::sampling}
\end{algorithm2e}
\DoubleSpacedXI

To analyze the revenue of Algorithm~\ref{alg::sampling}, we consider a hypothetical algorithm which behaves identically to Algorithm~\ref{alg::sampling}, except even when it defaults to line~\ref{line::default}, it is able to behave as if the sampling succeeded and makes the same decisions as lines~\ref{line::hit_start}--\ref{line::hit_end}.  Such an algorithm is equivalent to $\Exp$, and hereafter we will refer to it as $\Exp$.  The results of the sample runs do not affect the outcome of the algorithm, but help with bookkeeping.

\begin{definition}
Let $\fsamp_t$ be the indicator random variable for the sampling in Algorithm~\ref{alg::sampling} failing at time $t$, defined for all $t\in[T+1]$.  Let $\fsamp_{T+1}=1$ deterministically.  Analogously, let $\fexp_t$ be the indicator random variable for the sampling in $\Exp$ ``failing'' at time $t$, $\forall t\in[T+1]$.
\end{definition}

For convenience, here we will use different random variables to denote the valuations in the runs of $\Samp$ and $\Exp$: $\vsamp_t$ and $\vexp_t$, respectively.  We will also use the notation from Definition~\ref{def::diff_algs}.

\begin{definition}
Define the \emph{history up to time} $t$ to consist of realizations up to and including the sampling at time $t$.  Formally, for all $t\in[T+1]$, let
%	$\vh_t$ denote the vector
$\vh_t=(f_1,p_1,u_1,\ldots,f_{t-1},p_{t-1},u_{t-1},f_t)$, where:
\begin{itemize}
\item $f_{t'}$ is a binary variable in $\{0,1\}$ indicating whether the sampling failed at time $t'$, for all $t'\in[t]$;
\item $p_{t'}$ is a price in $\{r^{(1)},\ldots,r^{(m)},r^{(m+1)}\}$, for all $t'\in[t-1]$;
\item $u_{t'}$ is a valuation in $\{r^{(0)},r^{(1)},\ldots,r^{(m)}\}$, for all $t\in[t-1]$.
\end{itemize}
Furthermore, define the following vectors of random variables for all $t\in[T+1]$:
\begin{itemize}
\item $\vH^{\Samp}_t=(\fsamp_1,\psamp_1,\vsamp_1,\ldots,\fsamp_{t-1},\psamp_{t-1},\vsamp_{t-1},\fsamp_t)$;
\item $\vH^{\Exp}_t=(\fexp_1,\pexp_1,\vexp_1,\ldots,\fexp_{t-1},\pexp_{t-1},\vexp_{t-1},\fexp_t)$.
\end{itemize}
\end{definition}

Now, we would like to partition the sample paths by the history up to the first point of failure, and prove that the two algorithms behave identically up to this point.

\begin{definition}\label{def::failure_history}
Let $\cF_t$ denote the histories up to time $t$ such that the first failure in the sampling occurs at time $t$.  Formally, for all $t\in[T+1]$, $\cF_t$ is the set of $\vh_t=(f_1,p_1,u_1,\ldots,f_{t-1},p_{t-1},u_{t-1},f_t)$ such that $f_1=\ldots=f_{t-1}=0$ and $f_t=1$. ($p_1,\ldots,p_{t-1}$ and $u_1,\ldots,u_{t-1}$ are arbitrary, and thus $|\cF_t|=(m+1)^{2(t-1)}$.)
\end{definition}

\begin{lemma}\label{lem::coupling}
For a run of Algorithm~\ref{alg::sampling}, $\bigcup_{t=1}^{T+1}\bigcup_{\vh_t\in\cF_t}\{\vH^{\Samp}_t=\vh_t\}$ is a set of mutually exclusive and collectively exhaustive events.  Analogously, for a run of $\Exp$, $\bigcup_{t=1}^{T+1}\bigcup_{\vh_t\in\cF_t}\{\vH^{\Exp}_t=\vh_t\}$ is a set of mutually exclusive and collectively exhaustive events.

Furthermore, $\Pr[\vH^{\Samp}_t=\vh_t]=\Pr[\vH^{\Exp}_t=\vh_t]$, for all $t\in[T+1]$ and $\vh_t\in\cF_t$.
\end{lemma}

Lemma~\ref{lem::coupling} is straight-forward, so we defer its proof to E-supplement~\ref{sec::stochasticPf}.  Having proved it, we can write:
\begin{eqnarray}
\bE[\ALG^{\Samp}] & = & \sum_{t=1}^{T+1}\sum_{\vh_t\in\cF_t}\bE[\ALG^{\Samp}|\vH^{\Samp}_t=\vh_t]\Pr[\vH^{\Samp}_t=\vh_t] \label{eqn::couple_alg} \\
\bE[\ALG^{\Exp}] & = & \sum_{t=1}^{T+1}\sum_{\vh_t\in\cF_t}\bE[\ALG^{\Exp}|\vH^{\Exp}_t=\vh_t]\Pr[\vH^{\Exp}_t=\vh_t]. \label{eqn::couple_opt}
\end{eqnarray}
Since we also know that $\Pr[\vH^{\Samp}_t=\vh_t]=\Pr[\vH^{\Exp}_t=\vh_t]$, our goal is to compare the expected revenues of the two algorithms conditional on each history $\vh_t\in\cF_t$.

When $t=T+1$, i.e.\ the sampling never fails, it is easy to see that the two revenues are equal.  Indeed, for any $\vh_{T+1}\in\cF_{T+1}$:
\begin{eqnarray}
\bE[\ALG^{\Samp}|\vH^{\Samp}_{T+1}=\vh_{T+1}] & = & \bE\Big[\sum_{t=1}^T\psamp_t\cdot\bI(\vsamp_t\ge\psamp_t)\Big|\vH^{\Samp}_{T+1}=\vh_{T+1}\Big] \nonumber \\
& = & \sum_{t=1}^Tp_t\cdot\bI(u_t\ge p_t) \nonumber \\
& = & \bE[\ALG^{\Exp}|\vH^{\Exp}_{T+1}=\vh_{T+1}]. \label{eqn::no_fail}
\end{eqnarray}

\begin{lemma}
Recall that $\bE[\OPT(V_1,\ldots,V_T)]$ is the expected value of the offline optimum with $V_1,\ldots,V_T$ drawn independently according to $\vu_1,\ldots,\vu_T$.  For $t\le T$ and $\vh_t\in\cF_t$,
\begin{equation}\label{eqn::fail}
\bE[\ALG^{\Exp}|\vH^{\Exp}_t=\vh_t]-\bE[\ALG^{\Samp}|\vH^{\Samp}_t=\vh_t]\le\bE[\OPT(V_1,\ldots,V_T)],
\end{equation}
\end{lemma}

Substituting (\ref{eqn::no_fail}), for $\vh_{T+1}\in\cF_{T+1}$, and (\ref{eqn::fail}), for $\vh_1,\ldots,\vh_T\in\cF_1,\ldots,\cF_T$, into (\ref{eqn::couple_alg}) and (\ref{eqn::couple_opt}), we conclude that
\begin{equation}\label{eqn::final_frontier}
\bE[\ALG^{\Exp}]-\bE[\ALG^{\Samp}]\le\bE[\OPT]\cdot\Big(\sum_{t=1}^T\sum_{\vh_t\in\cF_t}\Pr[\vH^{\Samp}_t=\vh_t]\Big).
\end{equation}
By Definition~\ref{def::failure_history}, the expression in parentheses is the total probability of the sampling failing at any point, before choosing the final price $\psamp_T$.  We bound the term for each $t\in[T]$ separately.  As $t$ increases, the number of samples increases, so the probability of failure decreases:
\begin{lemma}\label{lem::prob_failing}
For all $t\in[T]$, $\sum_{\vh_t\in\cF_t}\Pr[\vH^{\Samp}_t=\vh_t]\le\frac{1}{eCt^2}.$
\end{lemma}

%remark. Intuitively, Lemma~\ref{lem::prob_failing} is true because scenarios where the sampling is likely to fail are unlikely scenarios in the first place.

It now follows easily that the sampling algorithm is within $\ve$ of being optimally competitive.
\begin{theorem}\label{thm::poly}
For all $\ve>0$, if we set $C=\lceil\frac{6}{e\pi^2\ve}\rceil$ in line~\ref{line::define_C} of Algorithm~\ref{alg::sampling}, then it is $(\frac{1}{q}-\ve)$-competitive, and has runtime polynomial in $\frac{1}{\ve}$, $k$, $T$, and $m$.
\end{theorem}
Theorem~\ref{thm::poly} is straight-forward and proved in E-supplement~\ref{sec::stochasticPf}.
% remark. probably num samples doesn't need to be t^2, only needs to be t, but bound is so crude. bleh.

\section{Extensions} \label{sec::extensions}

All proofs from this section are deferred to E-supplement~\ref{sec::extensionsPf}.

\subsection{A Continuum of Prices}\label{sec::continuum}

In this section we show how to modify Algorithm~\ref{alg::valuationTrackingProc} for the setting where valuations lie in $\{0\}\cup[\rmin,\rmax]$.  We let $R=\rmax/\rmin$ and rescale the price interval to be $[1,R]$.
The competitive ratio obtained will be $\frac{1}{1+\ln R}$, recovering the competitive ratio from \citet{BQ09}.

Consider Algorithm~\ref{alg::cont}.  Now $\val[i]$ keeps track of the highest valuation assigned to inventory unit $i$ thus far, starting at 0.  It is easy to see that the distributions specified in lines~\ref{line::proper_1} and \ref{line::proper_2} are proper.

\SingleSpacedXI
\begin{algorithm2e}[t]
\SetAlgoNoLine
\KwIn{Customers $t=1,2,\ldots$ arriving online, with each valuation $V_t$ revealed after the price $P_t$ is chosen.}
\KwOut{For each customer $t$, a (possibly random) price $P_t$ for her.}
$\val[i]=0,\sold[i]=\false$ for $i=1,\ldots,k$\;
$t=1$\;
\While{customer $t$ arrives}{
$v=\min_{i'}\{\val[i']\}$\;
$i=\min\{i':\val[i']=v\}$\;
\eIf{$\sold[i]=\false$}{
\eIf{v=0}{
offer price 1 w.p.\ $\frac{1}{1+\ln R}$, and price $r$ w.p.\ $\frac{1}{r(1+\ln R)}$ for all $r\in(1,R]$\;\label{line::proper_1}
}{
offer price $r$ w.p.\ $\frac{1}{r(\ln R-\ln v)}$ for all $r\in(v,R]$\;\label{line::proper_2}
}
}{
reject the customer by choosing price $\infty$\;
}
observe valuation $V_t$ and purchase decision $X_t$\;
\If{$V_t>v$}{
$\val[i]=V_t$\;
\If{$X_t=1$}{
$\sold[i]=\true$\;
}
}
$t=t+1$\;
}
\caption{Valuation Tracking Procedure for a Continuous Interval of Feasible Prices}\label{alg::cont}
\end{algorithm2e}
\DoubleSpacedXI

To analyze the competitiveness of Algorithm~\ref{alg::cont}, we prove lemmas analogous to Lemmas~\ref{lem::ALG_depletion}--\ref{lem::ALG_revenue}.  We use the same notation as in Definition~\ref{def::discrete_analysis}, except instead of $\ell_{i,t}$ and $j_t$, we use $w_{i,t}$ to denote the value of $\val[i]$ at the end of time $t$, taking a value in $\{0\}\cup[1,R]$.

\begin{lemma} \label{lem::contPrice1}
At the end of each time step $t$, the probability that any inventory unit $i$ has been sold is 0 if $w_{i,t}=0$, and $\frac{1+\ln w_{i,t}}{1+\ln R}$ if $w_{i,t}\ge1$.  Formally, for all $t=0,\ldots,T$,
\begin{equation}\label{eqn::cont_depletion}
\bE[S_{i,t}]=\bI(w_{i,t}>0)\cdot\frac{1+\ln w_{i,t}}{1+\ln R},\text{ for }i\in[k].
\end{equation}
\end{lemma}

%Now we analyze the expected revenue of the algorithm, which is $\bE[\ALG]$, or $\sum_{t=1}^T\bE[P_tX_t]$.
%As argued earlier, there cannot be a sale in a time step $t$ where $j_t\le\ell_{i_t,t-1}$, so for these time steps $X_t=0$ and $\bE[P_tX_t]=0$.  The following lemma derives the value of $\bE[P_tX_t]$ when $j_t>\ell_{i_t,t-1}$.
\begin{lemma} \label{lem::contPrice2}
Suppose $V_t=w_{i_t,t}>w_{i_t,t-1}$ in a time step $t\in[T]$.  Then the expected revenue earned by the algorithm during time step $t$ is $\frac{1}{1+\ln R}(w_{i_t,t}-w_{i_t,t-1})$.
\end{lemma}

With these two lemmas, the rest of the proof follows E-supplement~\ref{sec::valuationTrackingProcPf}.  Indeed, Lemma~\ref{lem::ALG_tot_revenue} says that $\bE[\ALG]=\frac{1}{1+\ln R}\sum_{i=1}^kw_{i,T}$.  Meanwhile, Lemma~\ref{lem::OPT_revenue} says that $\OPT=\sum_{i=1}^kw_{i,T}$.  Therefore, $\frac{\bE[\ALG]}{\OPT}\ge\frac{1}{1+\ln R}$, and since $V_1,\ldots,V_T$ was arbitrary, Algorithm~\ref{alg::cont} is $\frac{1}{1+\ln R}$-competitive.

\subsection{No Information on Valuations}\label{sec::rfpp}
In this section we discuss whether it is possible for an online algorithm to be $\CR(\cP)$-competitive without any information (before or after, deterministic or distributional) on the valuations.

First we show that this is impossible for any online algorithm which price-skims independently, i.e.\ realizes its random price at each time step using an independent source of random bits.

\begin{proposition}\label{prop::no_info}
Suppose that either: (i) $m\ge2$ and valuations can be 0 (as usual); or (ii) $m\ge3$ and valuations cannot be 0.  (Recall that $m$ is the number of prices.)  Then for any online algorithm where each $P_t$ chosen independently based on the sales history $X_1,\ldots,X_{t-1}$, there exists a sequence $V_1,\ldots,V_T$ such that
\begin{align*}
\frac{\bE[\ALG(V_1,\ldots,V_T)]}{\OPT(V_1,\ldots,V_T)}<\CR(\cP).
\end{align*}
\end{proposition}

However, we show that it is possible to be $\CR(\cP)$-competitive if the online algorithm can price-skim in a ``coordinated'' fashion, with the same probabilities as in \citet{EM10}.
\begin{proposition}\label{prop::random_yes}
Consider the following random-fixed-price policy:
\begin{enumerate}
\item Initially, choose a random price $P$ which is equal to each $r^{(j)}$ with probability $q^{(j)}/q$;
\item Offer price $P$ as long as there is remaining inventory.
\end{enumerate}
This policy is $\CR(\cP)$-competitive.
\end{proposition}

It is known that correlated randomness is very powerful in the design of online algorithms (see, e.g., \citet{KVV90}, who derive an extremely elegant solution to the online matching problem using correlated randomness).
Indeed, we can use our policy from Proposition~\ref{prop::random_yes} under our previous models with more information on the valuations and still have a $\CR(\cP)$-competitive algorithm.
However, this is impractical for several reasons.
First, the fact that a single random price is fixed makes the algorithm have large variance in its performance.
%makes it impossible to make use of additional information that may be available on the valuations (this issue was also raised in \citet{EM10}).
Second, the random-fixed-price policy does not show how the price should evolve as inventory is depleted; namely, it does not satisfy the intuitive structural property in dynamic pricing that the price is greater if the remaining inventory is less as we showed in Section~\ref{sec::structural}.
In Section~\ref{sec::compExp}, we will test the performance of the random-fixed-price policy, and see that in fact it performs worse than booking limits (which does not have a theoretical guarantee).

\subsection{Personalized Revenue Management Model}\label{sec::personalized}
In this section we consider the personalized online revenue management setup introduced by \citet{GNR14}, where:
\begin{itemize}
\item the stochastic decision of each customer can be modeled accurately upon her arrival to the e-commerce platform (by using her characteristics);
\item however, the overall intensity and characteristics of customers to arrive over time is difficult to model (and treated as unknown/arbitrary).
\end{itemize}
This corresponds to the stochastic-valuation model in Section~\ref{sec::stochastic}, with the change that during each time step $t$, the distribution of each $V_t$ is first given, and then the algorithm can offer a \textit{personalized} price.
The public pricing algorithms from Section~\ref{sec::stochastic} can still be applied, and will be $\CR(\cP)$-competitive.
Furthermore, it is not possible to be better than $\CR(\cP)$-competitive even with this personalized information, as discussed in Section~\ref{sec::setupIntro}.

Nonetheless, in this section we specify how our online algorithms can exploit personalized information to strictly improve their decisions, while remaining $\CR(\cP)$-competitive.
Take any $\CR(\cP)$-competitive algorithm $\cA$ for the stochastic-valuation model (e.g.\ the algorithm $\Exp$ from Section~\ref{sec::exp_time}).
%, or a modification following Section~\ref{sec::structural} which never offers price $\infty$ to customers before stocking out).
For each time step $t$ and inventory level $k'>0$ such that $\Pr[I^{\cA}_{t-1}=k']>0$, consider the distribution for the price $P_t$ chosen by algorithm $\cA$ conditioned on $I^{\cA}_{t-1}=k'$ (this depends on the previously-observed valuation distributions $\vu_1,\ldots,\vu_{t-1}$).  Since now we also know the distribution $\vu_t$ of valuation $V_t$, we can compute the probability of algorithm $\cA$ making a sale during time $t$,
\begin{align}
\sum_{j=1}^m\Pr[P^{\cA}_t=r^{(j)}|I^{\cA}_{t-1}=k']\Pr[V_t\ge r^{(j)}], \label{eqn::consumptionRate}
\end{align}
as well as its expected revenue,
\begin{align}
\sum_{j=1}^mr^{(j)}\Pr[P^{\cA}_t=r^{(j)}|I^{\cA}_{t-1}=k']\Pr[V_t\ge r^{(j)}]. \label{eqn::revenueRate}
\end{align}
We can interpret (\ref{eqn::revenueRate}) as the reward given to the algorithm during time $t$ in exchange for the probability (\ref{eqn::consumptionRate}) of consuming inventory.  The price distribution used by the algorithm to obtain such an exchange was chosen without knowing the distribution of $V_t$.  However, since now we do know the distribution of $V_t$, we can potentially make a decision which achieves \textit{more} expected reward under the same consumption probability.  Specifically, we solve the following LP:
\begin{align}
\max\sum_{j=1}^mr^{(j)}\Pr[V_t\ge r^{(j)}]p_j(t,k') & & \label{eqn::lpStart} \\
\text{s.t.}\sum_{j=1}^m\Pr[V_t\ge r^{(j)}]p_j(t,k') &=\sum_{j=1}^m\Pr[P^{\cA}_t=r^{(j)}|I^{\cA}_{t-1}=k']\Pr[V_t\ge r^{(j)}] & \label{eqn::consumptionMatching} \\
\sum_{j=1}^mp_j(t,k') &\le1 & \\
p_j(t,k') &\ge0 &\forall\ j=1,\ldots,m \label{eqn::lpEnd}
\end{align}
$p_j(t,k')$ represents the probability that we should offer price $j$ at time $t$, conditioned on the remaining inventory being $k'$.  We know that setting each $p_j(t,k')=\Pr[P^{\cA}_t=r^{(j)}|I^{\cA}_{t-1}=k']$ is a feasible solution, and hence the optimal objective value of the optimization problem is at least (\ref{eqn::revenueRate}).  Let $\{p^*_j(t,k'):j=1,\ldots,m\}$ denote an optimal solution to the optimization problem, for all $t$ and $k'$.

\begin{proposition}\label{prop::personalized}
Consider the online algorithm which, at each time step $t$, sets the price randomly according to probabilities $\{p^*_j(t,k'):j=1,\ldots,m\}$, where $k'$ is the remaining inventory at the start of time $t$.  Then for any sequence of valuation distributions $\vu_1,\ldots,\vu_T$, the total expected revenue of this algorithm is at least $\bE_{V_1\sim\vu_1,\ldots,V_T,\sim\vu_T}[\ALG^{\cA}(V_1,\ldots,V_T)]$.
\end{proposition}

Proposition~\ref{prop::personalized} is established in the same way as Theorems~\ref{thm::modifiedAlg} and \ref{thm::exp}---for $t=1,\ldots,T$, we can inductively ensure from constraint~(\ref{eqn::consumptionMatching}) that the distribution for the inventory level time $t$ is identical for both algorithms.  Since the modified algorithm has the same distribution for inventory state at each time $t$ and earns at least as much revenue as $\cA$ in expectation on every possible state, its total revenue can only be greater.  In fact, as we will see in Section~\ref{sec::compExp}, the modification with personalization often earns much greater revenue, since through solving the LP formed by~\eqref{eqn::lpStart}--\eqref{eqn::lpEnd}, it often finds a higher-revenue way to maintain the same inventory consumption probability.

Finally, we remark that if $\cP$ is a continuum $[\rmin,\rmax]$ (following the extension from Section~\ref{sec::continuum}), then the LP~\eqref{eqn::lpStart}--\eqref{eqn::lpEnd} would have infinitely many variables, but because it has only two constraints, an optimal solution would need support size at most two.
In fact, most parametric forms commonly used to describe a demand function over $[\rmin,\rmax]$, including linear, log-linear, and logit, are \textit{regular}, in which case an optimal solution has support of size one and the LP can be tractably solved \citep[cf.][Sec.~7.3]{TvR06}.

\section{Computational Experiments} \label{sec::compExp}

In this section we test the performance of our algorithm in simulations, following the personalized online revenue management setup (with stochastic valuations) discussed in Section~\ref{sec::personalized}.

\subsection{Experimental Setup and Algorithms Compared} \label{sec::compSetup}

The only information given initially is the starting inventory $k$ and the price set $\cP=\{r^{(1)},\ldots,r^{(m)}\}$, where we let $0=r^{(0)}<\ldots<r^{(m)}<r^{(m+1)}=\infty$.
Based on $k$ and $\cP$, we generate different arrival sequences $(\vu_1,\ldots,\vu_T)$ of varying lengths $T$.
The valuation distribution $\vu_t=(v^{(0)}_t,\ldots,v^{(m)}_t)$ for any particular customer $t=1,\ldots,T$ is log-linear (exponential).
That is, there is a price sensitivity parameter $b_t>0$ such that the probability of her valuation $V_t$ exceeding any price $P\ge0$ satisfies
\begin{align} \label{eqn::loglinear}
\Pr[V_t\ge P]=e^{-b_tP}.
\end{align}
This then induces a discrete distribution over the maximum-willingness-to-pay values of $r^{(0)},\ldots,r^{(m)}$ by setting $v^{(j)}_t=\Pr[V_t\ge r^{(j)}]-\Pr[V_t\ge r^{(j+1)}]$ for all $j=0,\ldots,m$; note that these probabilities $v^{(0)}_t,\ldots,v^{(m)}_t$ indeed sum to 1, by equation~\eqref{eqn::loglinear}.

We choose this family of distributions due to the convenience of equation~\eqref{eqn::loglinear} and the property that the immediate revenue $Pe^{-b_tP}$ is decreasing in $P$ whenever $P\ge1/b_t$ \citep[cf.][Sec.~7.3]{TvR06}.
This property is important for creating a non-trivial tradeoff between offering lower prices which maximize immediate revenue, vs.\ offering higher prices which maximize revenue per inventory.
We do not believe our results to be sensitive to the exact family of demand used as long as this property is satisfied.

For each arrival sequence we generate, we draw the parameters $b_1,\ldots,b_T$ independently from the same distribution, i.e.\ fluctuations in price sensitivity are stationary over time.
Therefore, arrival sequences are mainly distinguished by their total number of customers $T$,
where under small $T$ it is optimal to offer lower prices, while under large $T$ it is optimal to offer higher prices.
Of course, $T$ is initially unknown, and online algorithms must immediately decide a personalized price for each customer $t$ upon seeing her valuation distribution $\vu_t$ (equivalently, price sensitivity $b_t$).

We now describe the algorithms and benchmarks we test.
First we describe algorithms which are oblivious to both valuation distributions and remaining inventory.
\begin{itemize}
\item \textbf{PS} (Price-Skimming): choose a random price, according to \citet{EM10}, and fix that to be the price for all customers.
This policy is described in Section~\ref{sec::rfpp}.
\item \textbf{IPS} (Independent Price-Skimming): same as PS, but independently reset the random price for each customer.
This policy is described in Section~\ref{sec::attempts} (see ``Attempt 2'').
\item \textbf{Conservative}: always charge the maximum price of $r^{(m)}$.
\end{itemize}
Now we describe algorithms which are valuation-oblivious but react to remaining inventory.
\begin{itemize}
\item \textbf{BL} (Booking Limits):
gradually raise the prices as inventory is sold, according to \citet{BQ09}.
This policy is described in Section~\ref{sec::attempts} (see ``Attempt 1'').
\item \textbf{BL-PS} (Booking Limits with Price-Skimming):
same as BL, but instead of deterministically charging the base price, use PS to randomly choose a price above the base price.
This policy is described in Section~\ref{sec::attempts} (see ``Attempt 3'').
\end{itemize}

Now we modify the previous algorithms to offer personalized prices based on knowledge of valuation distributions.
Whenever the previous algorithm would have offered a base price, the modified algorithm looks for an opportunity to offer a higher price which also has a higher immediate revenue, for each customer $t$ (as discussed earlier, this could only be possible if the base price $P$ satisfies $P<1/b_t$).
In this way, we get the personalized modifications \textbf{PS-P}, \textbf{IPS-P}, \textbf{BL-P} of the earlier price-skimming, independent price-skimming, and booking limits methods for setting base prices, respectively.
BL-P reacts to remaining inventory but PS-P and IPS-P are inventory-oblivious.
Another similar inventory-oblivious algorithm we will test is the \textbf{Myopic} algorithm, which greedily offers to each customer $t$ the price $P$ among all prices which maximizes immediate revenue $Pe^{-b_tP}$, essentially doing personalization with a fixed base price of 1.

Finally, we test our \textbf{Valuation Tracking} procedure with stochastic valuations and personalization, as described in Section~\ref{sec::personalized}.
To be precise, we consider our public pricing algorithm $\Exp$ for stochastic valuations, sample 1000 times (see Section~\ref{sec::poly_time}) to estimate the price it would offer (replacing $\infty$-prices with $r^{(m)}$ as in Section~\ref{sec::structural}), and then look for an opportunity to offer a personalized price, repeating this process at each time step.
Our algorithm is similar to BL-P in that it is both valuation-aware and inventory-aware, but reacts to inventory in a different way.

For reference, we will also compare with the optimal \textbf{DP} (dynamic programming) benchmark which knows the entire arrival sequence $(\vu_1,\ldots,\vu_T)$ in advance (but still does not know the realizations of the valuations like $\OPT$ does), for a total of 11 algorithms tested.

\subsection{Parameter Values and Results} \label{sec::compResults}

We fix the price set to be $\cP=\{1,2,3,4\}$ for now, and vary its granularity and range later in Section~\ref{sec::simExt}.
We consider two scenarios for starting inventory: small ($k=10$) and large ($k=100$).
Under both values of $k$, we generate arrival sequences for 10 different lengths: $T=k,2k,\ldots,10k$.
We independently generate 1000 sequences for each value of $T$, resulting in a total of 10000 arrival sequence instances for each inventory scenario.

Recall from Section~\ref{sec::compSetup} that valuation distributions are log-linear.
We independently draw the price sensitivity parameter $b_t$ of each customer $t$ uniformly from the range [1/3,4/3], which ensures that even under the smallest value of $b_t=1/3$, the maximum price of 4 does not also maximize the immediate revenue function (since the immediate revenue $Pe^{-b_tP}$ decreases over prices $P\ge1/b_t$).
Meanwhile, the upper bound of $b_t=4/3$ was calibrated so that at the maximum length of $T=10k$, the Conservative algorithm can expect to sell most of the $k$ units of inventory at the maximum price (and hence is best-performing at the largest values of $T$, as will be evidenced).

For each instance, and each of the 11 algorithms from Section~\ref{sec::compSetup} as well as $\OPT$, we run 1000 simulations to empirically estimate the revenue earned (with the exception of DP, for which we can exactly calculate the expected revenue).
We divide the revenue earned by $\OPT$, to get a performance ratio for each algorithm on each instance.
We display the average performance of each algorithm over the 10000 instances, for both inventory scenarios $k=10$ and $k=100$, in Table~\ref{tbl::main}.

\begin{table}
\TABLE
{Average performance of each algorithm and benchmark over the instances, under both scenarios of starting inventory $k$.  The algorithms are sorted by whether they are valuation-aware and/or inventory-aware.  The performance of our algorithm is \textbf{bolded}.
\label{tbl::main}}
{
\begin{tabular}{|l|c|c|c|c|c|c|c|c|c|c|c|}
\hline
\updown & \multicolumn{4}{c|}{Valuation-oblivious} & \multicolumn{4}{c|}{Valuation-aware} & \multicolumn{3}{c|}{\multirow{2}{*}{Benchmarks}} \\
\updown & \multicolumn{2}{c|}{Inventory-obl.} & \multicolumn{2}{c|}{Inventory-aware} & \multicolumn{2}{c|}{Inventory-obl.} & \multicolumn{2}{c|}{Inventory-aware} & \multicolumn{3}{c|}{} \\
\updown & PS & IPS & BL & BL-PS & PS-P & IPS-P & BL-P & Val. Tracking & Myopic & Conservative & DP \\
\hline
\up $k=10$&48.0\%&45.8\%&55.5\%&57.9\%&54.3\%&54.5\%&61.3\%&\textbf{62.6}\%&49.3\%&49.3\%&73.7\% \\
\down $k=100$&47.9\%&45.6\%&56.6\%&59.2\%&54.3\%&54.5\%&62.4\%&\textbf{64.5}\%&49.1\%&48.7\%&76.1\% \\
\hline
\end{tabular}}
{}
\end{table}

We observe that there is noticeable separation in performance between the four categories of algorithms (valuation-oblivious vs.\ -aware, inventory-oblivious vs.\ -aware).  
Among the algorithms aware of neither valuations nor inventory, PS attains exactly the theoretically guarantee of $\CR(\{1,2,3,4\})=0.48$, while IPS performs worse.
Among the algorithms aware of both, Valuation Tracking outperforms BL-P.
All in all, awareness of inventory appears to be more important than awareness of valuations, especially at the larger starting inventory of $k=100$, and the three best algorithms---Valuation Tracking, BL-P, and BL-PS (which performs well relative to its category)---all raise their prices as the remaining inventory dwindles.
The intuition for why this yields good average-case performance over all the instances is that:
\begin{itemize}
\item When $T$ turns out to be small, the algorithms would have set low prices and maximized sales;
\item When $T$ turns out to be large, the algorithms would have raised their prices to get more out of their inventories.
\end{itemize}

\begin{figure}
\includegraphics[width=\textwidth]{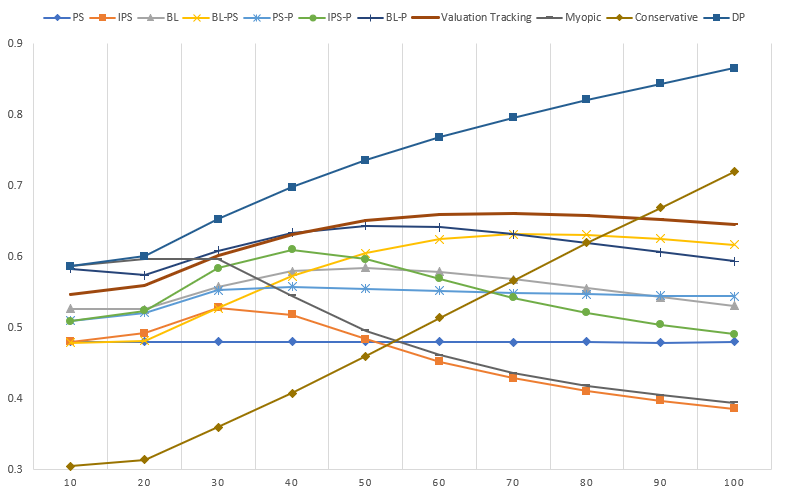}
\caption{Average performances of algorithms as the length $T$ varies over $10,20,\ldots,100$, with the starting inventory $k$ fixed at $10$.
The performance of our Valuation Tracking algorithm is \textbf{bolded}.}
\label{fig::compInvK10}
\end{figure}

\begin{figure}
\includegraphics[width=\textwidth]{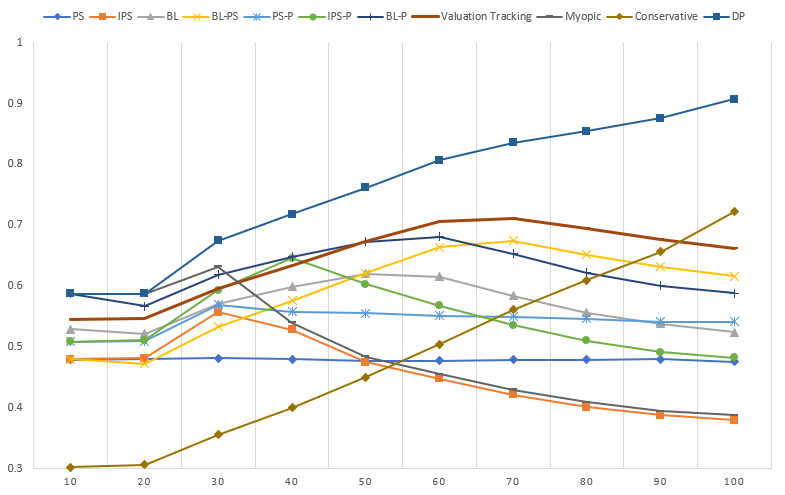}
\caption{Average performances of algorithms as the length $T$ varies over $100,200,\ldots,1000$, with the starting inventory $k$ fixed at $100$.
The performance of our Valuation Tracking algorithm is \textbf{bolded}.}
\label{fig::compInvK100}
\end{figure}

We now delve into why Valuation Tracking is better for reacting to inventory than booking limits.
We dis-aggregate the results from Table~\ref{tbl::main} and plot the performances of each algorithm as a function of $T$, averaging over the 1000 instances with that length.
We plot the results in Figures~\ref{fig::compInvK10}--\ref{fig::compInvK100}.

We observe the following trends in both Figure~\ref{fig::compInvK10} (for $k=10$) and Figure~\ref{fig::compInvK100} (for $k=100$).
\begin{itemize}
\item At the smallest and largest values of $T$, the Myopic and Conservative algorithms, which maximize revenue-per-customer and revenue-per-inventory respectively, are best-performing (aside from the future-knowing DP algorithm).
This further corroborates the legitimacy of our range of $T$ tested---not only are the aggregate performances of the Myopic and Conservative benchmarks comparable in Table~\ref{tbl::main}, we have covered the full spectrum of lengths from Myopic being best to Conservative being best.
\item The improvement of Valuation Tracking over BL-P can be attributed to noticeably better performance at large $T$, at the expense of slightly worse performance at small $T$.
\end{itemize}
The second bullet suggests that our algorithm is more conservative with inventory than booking limits.
To understand why, we have to look at the definitions of the algorithms.
Both algorithms essentially set aside the same fractions of inventory, given by $q^{(1)},\ldots,q^{(m)}$ from Definition~\ref{def::q}, ``to be sold at each price''.
However, booking limits are based on \textit{sales}, and offers low prices (assuming low prices maximize revenue-per-customer) at the start until a fixed amount (specifically, $q^{(1)}k$ units) of inventory is sold.
By contrast, our algorithm is based on the possible realizations of the \textit{valuation}, and starts considering higher prices from the start based on the exact valuation distribution.
%, and stops offering low prices once it has seen sufficiently many customers in expectation who could have had a high valuation.
%On long arrival sequences, this leads to fewer sales at lower prices, since BL-P will keep attempting lower prices until it realizes a fixed amount of sales, whereas Valuation Tracking will stop once it has made enough attempts based on expectation.

We also emphasize that Valuation Tracking cannot be matched by simply taking BL-P and making it more inventory-conservative by combining it with a more conservative algorithm.
Indeed, as evidenced in Figures~\ref{fig::compInvK10}--\ref{fig::compInvK100}, the performance curve of Valuation Tracking lies well above any convex combination formed by the other curves.
We believe that Valuation Tracking is fundamentally the more precise way to react to inventory under the stochastic nature of our pricing problem, while booking limits were invented for the revenue management problem where the decision is to deterministically accept/reject.

\subsection{Varying the Price Set} \label{sec::simExt}

We verify the robustness of our results from Section~\ref{sec::compResults} by varying the price set $\cP$ to be different from $\{1,2,3,4\}$.
We consider two ways to vary $\cP$:
\begin{itemize}
\item Increase the granularity of $\cP$ by allowing feasible prices to come in increments of 0.5 (i.e.\ $\cP=\{1.0,1.5,2.0,2.5,3.0,3.5,4.0\}$), 0.25, 0.125, etc., all the way down to increments of 0.015625.
% (in which case $|\cP|=193$).
\item Consider a maximum price of 2 or 8 instead, while maintaining granularity at the integer level (i.e.\ $\cP=\{1,2\}$ or $\cP=\{1,2,3,4,5,6,7,8\}$).
\end{itemize}

\begin{figure}
\includegraphics[width=\textwidth]{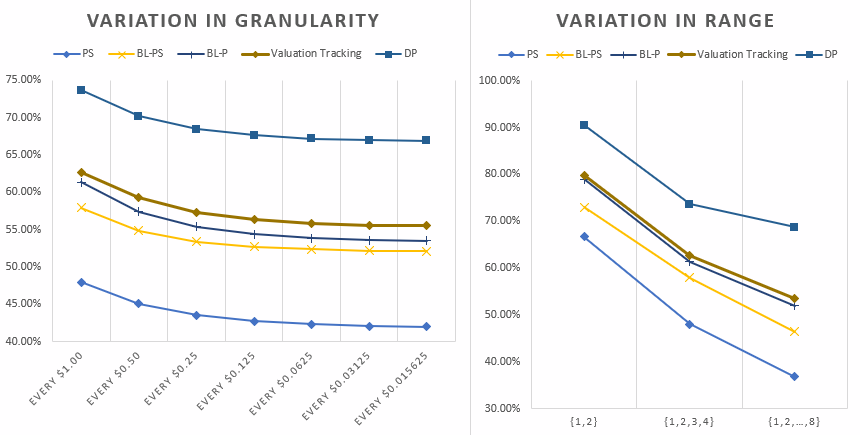}
\caption{Average performances of algorithms with $k$ fixed at 10 but $\cP$ allowed to vary.  Note that the performance of PS indicates the value of the competitive ratio guarantee $\CR(\cP)$.}
\label{fig::compVary}
\end{figure}

We repeat the experiments from Section~\ref{sec::compResults} for the small-inventory scenario of $k=10$, and plot the average performances of the algorithms over the instances in Figure~\ref{fig::compVary}.
For brevity, we only display the performances of PS (which always equals the competitive ratio guarantee of $\CR(\cP)$), the three best algorithms from Section~\ref{sec::compResults} (BL-PS, BL-P, and Valuation Tracking), and DP.

The relative performances of the algorithms in Figure~\ref{fig::compVary} does not differ from those originally observed in Table~\ref{tbl::main}.
In fact, a higher granularity of prices increases the advantage of our Valuation Tracking algorithm over BL-P, since with more possible pricing decisions there is more room for improvement.
Similarly, a larger range of prices makes valuation-awareness more important, and hence we see the drop in the performance of BL-PS when $\cP=\{1,2,\ldots,8\}$.

\section{Conclusion}\label{sec::conc}

In this paper we have studied the fundamental single-item dynamic pricing problem with no knowledge of future valuations, and derived the best-possible competitive ratio.
Our policies unify the inventory-dependent booking policies in \citet{BQ09} with the random price-skimming policies in \citet{EM10}.
An important feature of our policies is that they show at each time step how the \textit{price distribution} should depend on inventory when the future is \textit{unknown},
complementing classical results which show how the \textit{optimal price} should depend on inventory when the future is \textit{known}.
Our policies were derived using a new ``valuation tracking'' technique, which geometrically tracks the optimum and hedges against the arrival sequence immediately ending \textit{in the most inventory-conservative fashion}.
%We believe this to be of general interest for competitive ratio analysis.

Finally, we explain why our analysis of single-leg revenue management for dynamic pricing, where each customer has a valuation and chooses the lowest fare not exceeding it, captures substitution under any form of \textit{random-utility} choice model.
Suppose instead that the firm could offer an \textit{assortment} of fare classes, and that each customer has a \textit{ranked list} (in order of decreasing utility) of fare classes she is willing to purchase, and chooses the highest-ranked fare class that is offered to her.
We can define $V_t$ to be the maximum fare in the list that customer $t$ is willing to purchase, and then the offline optimum would still be the sum of the $k$ largest values from $V_1,\ldots,V_T$.
Meanwhile, we can modify the online algorithm so that whenever it would have offered price $P_t$, it now shows all fares greater than or equal to $P_t$.
This algorithm would still make a sale whenever $V_t\ge P_t$, except now it has the opportunity to earn revenue greater than $P_t$, if customer $t$ does not  choose the lowest offered fare.
As a result, our $\CR(\cP)$-competitive algorithms under the pricing model imply corresponding $\CR(\cP)$-competitive algorithms under the assortment model.

Nevertheless, we would like to end on two open questions related to the assortment generalization.  First, our argument above assumes random-utility choice models; however in practice certain fare classes could be designed as ``decoys'' for other fare classes. It is not known whether this effect can be accommodated by our dynamic pricing algorithm.
Second, our algorithms imply an ``assortment-skimming'' distribution over revenue-ordered assortments, but this assumes there is no limit on the number of fare classes offered.  We believe that assortment skimming under cardinality constraints is an interesting problem.

\section*{Acknowledgments}
The authors would like to thank He Wang for insightful discussions.

\bibliographystyle{informs2014} % outcomment this and next line in Case 1
\bibliography{bibliography} % if more than one, comma separated

%% Here starts the e-companion (EC)
%%%%%%%%%%%%%%%%%%%%%%%%%%%%%%%%%%%%%%%%%%%%%%%%%%%%%%%%%%
\ECSwitch

%\ECDisclaimer
%%%%%%%%%%%%%%%%%%%%%%%%%%%%%%%%%%%%%%%%%%%%%%%%%%%%%%%%%%

%%% Main head for the e-companion
\ECHead{E-Companion}

\begin{APPENDICES}

\section{Why Attempts using Existing Techniques Fail even with Personalization}\label{sec::newEg}

We modify the example from Attempt~3 of Section~\ref{sec::attempts} to show that a personalized adaptation of booking limits would still fail to be $\CR(\cP)$-competitive under stochastic valuations.
Recall that we are using price set $\cP=\{1,2,4\}$ and starting inventory $k=4$.
Suppose that the first two customers have a valuation which is 4 w.p.\ $1/4-\ve$, 2 w.p.\ $1/4-\ve$, and 1 w.p.\ $1/2+2\ve$, for some small $\ve>0$.
Since the booking limit would always be 1 for the first two customers, a personalized adaptation of the policy from \citet{BQ09} would still offer them the price of 1, which maximizes immediate revenue.
Afterward, the booking limit would be raised to 2, in which case a third customer whose valuation is deterministically 1 would be rejected.  The total revenue earned is 2.
However, for this example, the expected value of the optimum is $\approx2+2+1=5$.
Therefore, the desired competitiveness of $\CR(\{1,2,4\})=1/2$ is not achieved.

\section{Full Proof of Theorem~\ref{thm::valuationTrackingProc}}\label{sec::valuationTrackingProcPf}

\begin{definition}\label{def::discrete_analysis}
Define the following:
\begin{itemize}
\item $S_{i,t}$: the indicator random variable for whether inventory unit $i$ is sold by the end of time $t$, i.e.\ the value of $\sold[i]$ at the end of time $t$, defined for all $i\in[k]$ and $t=0,\ldots,T$;
\item $i_t$: the inventory unit assigned to customer $t$, taking a value in $[k]$ for all $t\in[T]$;
\item $\ell_{i,t}$: the value such that $\level[i]=r^{(\ell_{i,t})}$ at the end of time $t$, taking a value in $\{0,1,\ldots,m\}$ for all $i\in[k]$ and $t=0,\ldots,T$;
\item $j_t$: the value in $\{0,1,\ldots,m\}$ such that $V_t=r^{(j_t)}$, defined for all $t\in[T]$.
\end{itemize}
\end{definition}

Fix the deterministic sequence of valuations $V_1,\ldots,V_T$ chosen by the adversary.  $i_t$, $\ell_{i,t}$, and $j_t$ are not random variables; they are determined by $V_1,\ldots,V_T$.

We would like to write the random variables $S_{i,t}$ in terms of the other random variables.  By definition, $S_{i,0}=0$ for all $i\in[k]$.  For $t>0$, the following equations hold:
\begin{eqnarray}
S_{i_t,t} & = & S_{i_t,t-1}+X_t; \label{eqn::basic_S_X} \\
S_{i,t} & = & S_{i,t-1},\text{ for }i\neq i_t. \label{eqn::basic_S_X_2}
\end{eqnarray}

(\ref{eqn::basic_S_X})--(\ref{eqn::basic_S_X_2}) are easy to see.  In the algorithm, the only inventory unit that could potentially be sold during time $t$ is $i_t$.  This explains why (\ref{eqn::basic_S_X_2}) holds for all $i\neq i_t$.  It also explains why $S_{i_t,t}=1$ if and only if $S_{i_t,t-1}=1$ or $X_t=1$.  Furthermore, $S_{i_t,t-1}$ and $X_t$ cannot both be 1, since the algorithm does not try to sell inventory unit $i_t$ again at time $t$ if it has already been sold.  This completes the explanation for (\ref{eqn::basic_S_X}).

We now analyze the state of the $\sold$ array during the execution of the algorithm.
\begin{lemma}\label{lem::ALG_depletion}
At the end of each time step $t$, the probability that any inventory unit $i$ has been sold is $\frac{1}{q}\sum_{j=1}^{\ell_{i,t}}q^{(j)}$.  Formally, for all $t=0,\ldots,T$,
\begin{equation}\label{eqn::ALG_depletion}
\bE[S_{i,t}]=\frac{1}{q}\sum_{j=1}^{\ell_{i,t}}q^{(j)},\text{ for }i\in[k].
\end{equation}
\end{lemma}
\proof{Proof.}
We proceed by induction on $t$.  (\ref{eqn::ALG_depletion}) is true at time $t=0$, where $\bE[S_{i,0}]=0$ and $\ell_{i,0}=0$ for all $i\in[k]$.

Now suppose we are at the end of some time $t>0$ and (\ref{eqn::ALG_depletion}) was true at the end of time $t-1$.  We need to prove that (\ref{eqn::ALG_depletion}) is still true at the end of time $t$.  For $i\neq i_t$, $S_{i,t}=S_{i,t-1}$, by (\ref{eqn::basic_S_X_2}).  The value of $\level[i]$ is unchanged by the algorithm during time $t$, so $\ell_{i,t}=\ell_{i,t-1}$ as well.  The inductive hypothesis from time $t-1$ then establishes that $\bE[S_{i,t}]=\frac{1}{q}\sum_{j=1}^{\ell_{i,t}}q^{(j)}$.

It remains prove $\bE[S_{i_t,t}]=\frac{1}{q}\sum_{j=1}^{\ell_{i_t,t}}q^{(j)}$.  This is immediate if $j_t$ is no greater than $\ell_{i_t,t-1}$ (the value of the $\ell$ variable during iteration $t$ of the algorithm), since both $S_{i,t}$ and $\ell_{i,t}$ would be unchanged.  If $j_t>\ell_{i_t,t-1}$, the following can be derived (let $\ell=\ell_{i_t,t-1}$ for brevity):
\begin{eqnarray*}
\bE[S_{i_t,t}] & = & \bE[S_{i_t,t-1}]+\bE[X_t] \\
& = & \bE[S_{i_t,t-1}]+\bE[X_t|S_{i_t,t-1}=0]\cdot\Pr[S_{i_t,t-1}=0] \\
& = & \frac{1}{q}\sum_{j=1}^{\ell}q^{(j)}+\Pr[X_t=1|S_{i_t,t-1}=0]\Big(1-\frac{1}{q}\sum_{j=1}^{\ell}q^{(j)}\Big) \\
& = & \frac{1}{q}\sum_{j=1}^{\ell}q^{(j)}+\Big(\sum_{j=\ell+1}^{j_t}\frac{q^{(j)}}{\sum_{j'=\ell+1}^mq^{(j')}}\Big)\Big(\frac{\sum_{j=\ell+1}^{m}q^{(j)}}{q}\Big) \\
& = & \frac{1}{q}\sum_{j=1}^{j_t}q^{(j)}. \\
\end{eqnarray*}
The first equality follows from (\ref{eqn::basic_S_X}) and the linearity of expectation.  The second equality conditions on $S_{i_t,t-1}$ being 0, since the value of $X_t$ is 0 if $S_{i_t,t-1}=1$.  The third equality uses the value of $\bE[S_{i_t,t-1}]$ guaranteed by the inductive hypothesis.  In the fourth equality, the probability of getting a sale, conditioned on Algorithm~\ref{alg::valuationTrackingProc} reaching line~\ref{line::sum_to_one}, is equal to the probability of choosing a price at most $r^{(j_t)}$, the valuation of customer $t$.  The final equality achieves the desired result because $j_t=\ell_{i_t,t}$, the new value for $\level[i]$ after line~\ref{line::increment_anyway} of iteration $t$ of the algorithm.

This completes the induction and the proof of the lemma.
\Halmos\endproof

Now we analyze the expected revenue of the algorithm, which is $\bE[\ALG]$, or $\sum_{t=1}^T\bE[P_tX_t]$.  As argued earlier, there cannot be a sale in a time step $t$ where $j_t\le\ell_{i_t,t-1}$, so for these time steps $X_t=0$ and $\bE[P_tX_t]=0$.  The following lemma derives the value of $\bE[P_tX_t]$ when $j_t>\ell_{i_t,t-1}$.
\begin{lemma}\label{lem::ALG_revenue}
Suppose $j_t>\ell_{i_t,t-1}$ in a time step $t\in[T]$.  Then the expected revenue earned by the algorithm during time step $t$ is $\frac{1}{q}(r^{(j_t)}-r^{(\ell_{i_t,t-1})})$.
\end{lemma}
\proof{Proof.}
Let $t\in[T]$ be any time step for which $j_t>\ell_{i_t,t-1}$.  For brevity, let $\ell$ denote $\ell_{i_t,t-1}$.  The following can be derived:
\begin{eqnarray*}
\bE[P_tX_t] & = & \bE[P_tX_t|S_{i_t,t-1}=0]\Pr[S_{i_t,t-1}=0] \\
& = & \Big(\sum_{j=\ell+1}^mr^{(j)}\bE[X_t|P_t=r^{(j)}]\Pr[P_t=r^{(j)}|S_{i_t,t-1}=0]\Big)\Big(1-\Pr[S_{i_t,t-1}=1]\Big) \\
& = & \Big(\sum_{j=\ell+1}^mr^{(j)}\bI[j_t\ge j]\Pr[P_t=r^{(j)}|S_{i_t,t-1}=0]\Big)\Big(1-\frac{1}{q}\sum_{j'=1}^{\ell}q^{(j')}\Big) \\
& = & \Big(\sum_{j=\ell+1}^{j_t}r^{(j)}\frac{q^{(j)}}{\sum_{j'=\ell+1}^mq^{(j')}}\Big)\Big(\frac{\sum_{j'=\ell+1}^{m}q^{(j')}}{q}\Big) \\
& = & \frac{1}{q}\sum_{j=\ell+1}^{j_t}r^{(j)}(1-\frac{r^{(j-1)}}{r^{(j)}})
\end{eqnarray*}
The first equality conditions on $S_{i_t,t-1}$ being 0; note that $X_t=0$ if $S_{i_t,t-1}=1$.  The second equality conditions on the value of $P_t$, where we drop the conditioning on $S_{i_t,t-1}$ in the term $\bE[X_t|P_t=r^{(j)}]$ since $P_t\neq\infty$ already implies $S_{i_t,t-1}=0$.  This term becomes $\bI[j_t\ge j]$ in the third equality, since it is deterministically 1 or 0 depending on whether $V_t\ge r^{(j)}$, or equivalently $j_t\ge j$.  The third equality also uses Lemma~\ref{lem::ALG_depletion}, for the value of $\Pr[S_{i_t,t-1}=1]$.  The fourth equality uses the offering probabilities from line~\ref{line::sum_to_one} of Algorithm~\ref{alg::valuationTrackingProc}.  The fifth equality uses the explicit definition of $q^{(j)}$ from Definition~\ref{def::q}, and it is easy to see that the final expression is equal to $\frac{1}{q}(r^{(j_t)}-r^{(\ell)})$.
\Halmos\endproof

Lemma~\ref{lem::ALG_revenue} in turn implies the following lemma.
\begin{lemma}\label{lem::ALG_tot_revenue}
The expected revenue earned by the algorithm up to time $t$, $\sum_{t'=1}^t\bE[P_{t'}X_{t'}]$, is $\frac{1}{q}\sum_{i=1}^kr^{(\ell_{i,t})}$.
\end{lemma}
\proof{Proof.}
The customers up to time $t$ can be partitioned according to the inventory unit they were assigned, so
\begin{equation}\label{eqn::6845}
\sum_{t'=1}^t\bE[P_{t'}X_{t'}]=\sum_{i=1}^k\sum_{t'\le t:i_{t'}=i}\bE[P_{t'}X_{t'}].
\end{equation}
Consider any $i$.  For each $t'$ assigned to $i$, $\bE[P_{t'}X_{t'}]$ is 0 if $j_{t'}\le\ell_{i,t'-1}$.  Denote the remaining $t'$ such that $j_{t'}>\ell_{i,t'-1}$ by $t'_1,\ldots,t'_N$, where $N\ge0$ and $t'_1<\ldots<t'_N$.  Using Lemma~\ref{lem::ALG_revenue},
\begin{eqnarray*}
\sum_{t'\le t:i_{t'}=i}\bE[P_{t'}X_{t'}] & = & \sum_{n=1}^N\frac{1}{q}(r^{(j_{t'_n})}-r^{(\ell_{i,t'_n-1})}). \\
\end{eqnarray*}
Before time $t'_n$, $\level[i]$ was last updated at time $t'_{n-1}$, so $\ell_{i,t'_n-1}=j_{t'_{n-1}}$.  Therefore, the sum telescopes and the remaining term is $\frac{1}{q}r^{(j_{t'_N})}$ (note that $r^{(\ell_{i,t'_1-1})}=r^{(0)}=0$).  Now, $j_{t'_N}=\ell_{i,t'_N}$, and $\level[i]$ is not updated again in time steps $t'_N+1,\ldots,t$, so $\ell_{i,t'_N}=\ell_{i,t}$.  Substituting $\sum_{t'\le t:i_{t'}=i}\bE[P_{t'}X_{t'}]=\frac{1}{q}r^{(\ell_{i,t})}$ into (\ref{eqn::6845}) completes the proof.
\Halmos\endproof

Having established the revenue of our online algorithm, we compare it to the offline optimum.  Knowing the sequence of valuations $V_1,\ldots,V_T$ in advance, it is clear that the following algorithm is optimal:
\begin{enumerate}
\item Find the $\min\{k,T\}$ customers with the largest valuations;
\item Charge each of these customers $t$ her maximum willingness-to-pay $V_t$;
\item Do not sell to any other customer.
\end{enumerate}
The revenue $\OPT$ would be the the sum of the $\min\{k,T\}$ largest valuations.

\begin{definition}\label{def::free_disposal}
For all $t\in[T]$, let $M^k(t)$ be a vector consisting of the $k$ largest elements from $(V_1,\ldots,V_t)$, in any order.  If $t<k$, fill in the remaining entries of $M^k(t)$ with zeros.
\end{definition}
Then $\OPT=\sum_{i=1}^kM^k_i(T)$, where $M^k_i(T)$ denotes the $i$'th entry of $M^k(T)$.  It turns out that $M^k(t)$ is closely tracked by the $\level$ array from Algorithm~\ref{alg::valuationTrackingProc}, as $t$ progresses from 1 to $T$.  Both $\ell_{i,t}$ (the value of $\level[i]$ at the end of time $t$) and $M^k(t)$ are deterministic functions of $V_1,\ldots,V_t$.

\begin{lemma}\label{lem::OPT_revenue}
For all $t=0,\ldots,T$, the entries of the vector $(r^{(\ell_{1,t})},\ldots,r^{(\ell_{k,t})})$ is a permutation of the entries of the vector $M^k(t)$.
\end{lemma}
\proof{Proof.}
We proceed by induction on $t$.  At time $t=0$, both $M^k(0)$ and $(r^{(\ell_{1,0})},\ldots,r^{(\ell_{k,0})})$ is a vector of $k$ zeros, so the statement is true.

Now consider $t>0$, and suppose that $M^k(t-1)$ is a permutation of $(r^{(\ell_{1,t-1})},\ldots,r^{(\ell_{k,t-1})})$.  Therefore, a minimum entry in $M^k(t-1)$ is equal to a minimum entry in $(r^{(\ell_{1,t-1})},\ldots,r^{(\ell_{k,t-1})})$, which in turn is equal to $r^{(\ell_{i_t,t-1})}$, by Definition~\ref{def::discrete_analysis}.

If $j_t>\ell_{i_t,t-1}$, or equivalently $V_t=r^{(j_t)}>r^{(\ell_{i_t,t-1})}$, then by the definition of $M^k(t)$, $V_t$ must be added to $M^k(t-1)$ and replace any minimum entry equal to $r^{(\ell_{i_t,t-1})}$.  Meanwhile, $\ell_{i_t,t}=j_t$, and $\ell_{i,t}=\ell_{i,t-1}$ for all $i\neq i_t$, thus the only change from $(r^{(\ell_{1,t-1})},\ldots,r^{(\ell_{k,t-1})})$ to $(r^{(\ell_{1,t})},\ldots,r^{(\ell_{k,t})})$ is that the entry at index $i_t$ has been replaced by $r^{(j_t)}$.  Since $M^k(t-1)$ and $(r^{(\ell_{1,t-1})},\ldots,r^{(\ell_{k,t-1})})$ go through the same change at time $t$, $M^k(t)$ is still a permutation of $(r^{(\ell_{1,t})},\ldots,r^{(\ell_{k,t})})$.

If instead $j_t\le\ell_{i_t,t-1}$, then every entry of $M^k(t-1)$ is already at least $r^{(j_t)}=V_t$, so $M^k(t-1)$ incurs no change at time $t$.  Similarly, $\ell_{i,t}=\ell_{i,t-1}$ for all $i\in[k]$, so $(r^{(\ell_{1,t-1})},\ldots,r^{(\ell_{k,t-1})})$ incurs no change as well.  In both cases, we have established that $M^k(t)$ is a permutation of $(r^{(\ell_{1,t})},\ldots,r^{(\ell_{k,t})})$, completing the induction and the proof.
\Halmos\endproof

With Lemma~\ref{lem::ALG_tot_revenue} and Lemma~\ref{lem::OPT_revenue}, it is easy to establish the competitiveness of Algorithm~\ref{alg::valuationTrackingProc}.  Our online algorithm is designed so that during each time step $t$, it earns exactly $\frac{1}{q}$ of the amount that the offline optimum would increase by with the addition of $V_t$, and it does not need to observe $V_t$ beforehand to accomplish this.

\proof{Proof of Theorem~\ref{thm::valuationTrackingProc}.}
Fix any sequence of valuations $(V_1,\ldots,V_T)$.  $\bE[\ALG]$ is equal to $\sum_{t=1}^T\bE[P_tX_t]$, which in turn is equal to $\frac{1}{q}\sum_{i=1}^kr^{(\ell_{i,T})}$, by Lemma~\ref{lem::ALG_tot_revenue}.  Meanwhile, $\OPT=\sum_{i=1}^kM^k_i(T)$, and the entries of $M^k(T)$ is a permutation of the entries of $(r^{(\ell_{1,T})},\ldots,r^{(\ell_{k,T})})$, by Lemma~\ref{lem::OPT_revenue}.  Therefore, $\OPT=\sum_{i=1}^kr^{(\ell_{i,T})}=q\cdot\bE[\ALG]$, completing the proof of Theorem~\ref{thm::valuationTrackingProc}.
\Halmos\endproof

\section{Proofs from Sections~\ref{sec::modifiedAlg}--\ref{sec::structural}}\label{sec::modifiedAlgPf}

\proof{Proof of Lemma~\ref{lem::polytime_dp}.}
Let $i=i^*_t$, for brevity.  Let $S_{i,t}$ be the indicator random variable for inventory unit $i$ being sold by the end of time $t$.  For all $k'\in\{0,\ldots,k\}$,
\begin{equation}\label{eqn::bayes}
\Pr[S_{i,t-1}=0|I_{t-1}=k']=\frac{\Pr[I_{t-1}=k'|S_{i,t-1}=0]\Pr[S_{i,t-1}=0]}{\Pr[I_{t-1}=k'|S_{i,t-1}=0]\Pr[S_{i,t-1}=0]+\Pr[I_{t-1}=k'|S_{i,t-1}=1]\Pr[S_{i,t-1}=1]}
\end{equation}
by Bayes' law.  For all $t\in[T]$, $i\in[k]$, and $k'\in\{0,\ldots,k\}$, we explain how to compute $\Pr[I_{t-1}=k'|S_{i,t-1}=0]$ in polynomial time; $\Pr[I_{t-1}=k'|S_{i,t-1}=1]$ can be computed analogously.

First we argue that the Bernoulli random variables $\{S_{i',t-1}:i'\in[k]\}$ are independent.  To see this, note that the assignment procedure in Algorithm~\ref{alg::valuationTrackingProc} is deterministic.  Therefore, each $S_{i',t-1}$ is only dependent on the prices chosen for the customers assigned to $i'$, and while these prices could be dependent on each other, they are independent from the prices chosen for customers not assigned to $i'$.

Furthermore, $I_{t-1}=k-\sum_{i'=1}^kS_{i',t-1}$.  By independence, $\Pr[I_{t-1}=k'|S_{i,t-1}=0]=\Pr[\sum_{i'\neq i}S_{i',t-1}=k-k']$.  $\sum_{i'\neq i}S_{i',t-1}$ is simply the sum of $k-1$ independent Bernoulli random variables with known mean (from Lemma~\ref{lem::ALG_depletion}), hence the probability that it equals a specific value can be computed using dynamic programming.

We elaborate on the dynamic programming.  For notational convenience, without loss of generality assume $i=k$.  We will inductively for $a=0,\ldots,k-1$ maintain the value of $\Pr[\sum_{i'=1}^aS_{i',t-1}=b]$ for all $b\in\{0,\ldots,k\}$.  It is easy to initialize this for $a=0$.  Given $\Pr[\sum_{i'=1}^aS_{i',t-1}=b]$ for all $b\in\{0,\ldots,k\}$, note that
\begin{equation*}
\Pr[\sum_{i'=1}^{a+1}S_{i',t-1}=b]=\Pr[\sum_{i'=1}^aS_{i',t-1}=b-1]\Pr[S_{a+1,t-1}=1]+\Pr[\sum_{i'=1}^aS_{i',t-1}=b]\Pr[S_{a+1,t-1}=0]
\end{equation*}
for all $b\in\{0,\ldots,k\}$.  Each iteration of $a$ can be computed in time linear in $k$, and there are less than $k$ iterations.

$\Pr[I_{t-1}=k'|S_{i,t-1}=1]$ can be computed analogously.  It is clear that both procedures can be done in time $O(k^2)$ (ignoring the $O(t)$ time it may take to compute the assignment procedure), completing the proof of Lemma~\ref{lem::polytime_dp}.
\Halmos\endproof

\proof{Proof of Lemma~\ref{lem::detVInd} and Theorem~\ref{thm::modifiedAlg}.}
We argue that Lemma~\ref{lem::detVInd} and Theorem~\ref{thm::modifiedAlg} are the special cases of Lemma~\ref{lem::exp_states} and Theorem~\ref{thm::exp} from the stochastic-valuation model.  It is easy to check that the statements are analogous, so it suffices to show that $\Exp$ (as defined in Section~\ref{sec::exp_time}) executed on deterministic valuations is identical to Algorithm~\ref{alg::valuationTrackingProc}' (as defined in Section~\ref{sec::modifiedAlg}).

We show that the decision rule for a single time period $t$, and any amount of inventory remaining $k'$, is the same.  Let $i=i^*_t$ and $\ell=\ell_t$, for brevity.  Consider the values of $i$ and $\ell$ during iteration $t$ of Algorithm~\ref{alg::valuationTrackingProc} (with the deterministic valuations $V_1,\ldots,V_T$).  First consider any $j=\ell+1,\ldots,m$.	
\begin{align*}
\Pr[\pvkp_t=r^{(j)}|I^{\VKp}_{t-1}=k'] &=\Pr[S_{i,t-1}=0|I^{\VK}_{t-1}=k']\cdot\frac{q^{(j)}}{\sum_{j'=\ell+1}^mq^{(j')}} \\
&=\Pr[S_{i,t-1}=0|I^{\VK}_{t-1}=k']\Pr[P^{\VK}_t=r^{(j)}|S_{i,t-1}=0,I^{\VK}_{t-1}=k'] \\
&=\Pr[P^{\VK}_t=r^{(j)}\cap S_{i,t-1}=0|I^{\VK}_{t-1}=k'] \\
&=\Pr[P^{\VK}_t=r^{(j)}|I^{\VK}_{t-1}=k']
\end{align*}
The first equality holds by the specification of algorithm Algorithm~\ref{alg::valuationTrackingProc}'.  The second equality holds by the specification of Algorithm~\ref{alg::valuationTrackingProc}, where we can add the conditioning on $I^{\VK}_{t-1}=k'$ in the second probability due to independence.  The final equality follows because $P^{\VK}_t=r^{(j)}\neq\infty$ implies $S_{i,t-1}=0$.

If $j=m+1$, then
\begin{align*}
\Pr[\pvkp_t=\infty|I^{\VKp}_{t-1}=k'] &=\Pr[S_{i,t-1}=1|I^{\VK}_{t-1}=k'] \\
&=\Pr[P^{\VK}_t=\infty|I^{\VK}_{t-1}=k']
\end{align*}
since the event $S_{i,t-1}=1$ occurs if and only if the event $P^{\VK}_t=\infty$ occurs.

Finally, clearly if $j\le\ell$, then both $\Pr[\pvkp_t=r^{(j)}|I^{\VKp}_{t-1}=k']$ and $\Pr[\pvk_t=r^{(j)}|I^{\VK}_{t-1}=k']$ are 0.

We have shown that $\Pr[\pvkp_t=r^{(j)}|I^{\VKp}_{t-1}=k']=\Pr[\pvk_t=r^{(j)}|I^{\VK}_{t-1}=k']$ for all $j\in\{1,\ldots,m,m+1\}$, so it is the same decision rule as $\Exp$, completing the proof.
\Halmos\endproof

\proof{Proof of Lemma~\ref{lem::strictDominance}.}
Fix a valuation sequence $V_1,\ldots,V_T$ and consider any sample path in the execution of $\cA$; let the sample path be depicted by the sequence of random prices $P^{\cA}_1,\ldots,P^{\cA}_T$.  The revenue $\ALG^{\cA}$ on that sample path is given by $\sum_{t:V_t\ge P^{\cA}_t}P^{\cA}_t$; note that the cardinality of the set $\{t:V_t\ge P^{\cA}_t\}$ is at most $k$.

On that same sample path, the modified algorithm $\cA'$ would sell to the $k$ customers with the smallest indices in $\{t:V_t\ge\min\{P^{\cA}_t,r^{(m)}\}\}$ (or all the customers in that set if its cardinality is less than $k$).  Let $\cS$ denote the set of customers served by the modified algorithm.  Let $b=|\{t\in\cS:P^{\cA}_t=\infty\}|$, the number of customers with valuation $r^{(j)}$ served by the modified algorithm that were not served by the original algorithm.

It is easy to see that
\begin{align}
\ALG^{\cA'}-\ALG^{\cA} &=\sum_{t\in\cS}\min\{P^{\cA}_t,r^{(m)}\}-\sum_{t:V_t\ge P^{\cA}_t}P^{\cA}_t \nonumber \\
&\ge br^{(m)}-\sum_{t\in\cS'}P^{\cA}_t \label{eqn::2142}
\end{align}
where $\cS'$ is the set of customers that are no longer served by $\cA'$ because it used up $b$ extra units of inventory.  Since $|\cS'|\le b$, and $P^{\cA}_t\le r^{(m)}$ for all $t$ such that $P^{\cA}_t\le V_t$, it is immediate that (\ref{eqn::2142}) is non-negative.  Since this holds on every sample path for $\cA$, we have completed the proof that $\bE[\ALG^{\cA'}]\ge\bE[\ALG^{\cA}]$.
\Halmos\endproof

\proof{Proof of Theorem~\ref{thm::structural}.}
The inventory level $I_{t-1}$ is equal to $k-\sum_{i=1}^k(1-S_{i,t-1})$, where $S_{i,t-1}$ is the indicator random variable for inventory unit $i$ being sold by the end of time $t-1$.  We will hereafter omit the subscript $t-1$.

Each term $(1-S_i)$ is independent and equal to 1 with probability $(\sum_{j=b_i+1}^m)q^{(j)}/q$, which is the probability that inventory unit $i$ has not been sold.  We will denote it using $p_i$ and let $Y_i=1-S_i$, for brevity.  As long as $b_i$ (the index in $0,\ldots,m$ of the highest valuation assigned to inventory unit $i$) is not 0 or $m$, $p_i\in(0,1)$.
We will without loss of generality assume that $p_i\in(0,1)$ for all $i$, redefining $k$ and re-indexing as necessary (if $p_i=0$ or $p_i=1$ then $Y_i$ is deterministic and we can remove it from analysis of the random sum).
By the assumptions in the statement of the theorem, this re-indexing does not cause $i^*_t$ to fall outside of $1,\ldots,k$; in fact we can without loss of generality assume $i^*_t=1$.
Furthermore, $k_1,k_2$ are at least 0 at at most the re-defined $k$, since they correspond to inventory levels that are realized with non-zero probability.

After all of these transformations, the
statement reduces to
%desired statement to prove is
\begin{align}
\Pr[Y_1=1|\sum_{i=1}^kY_i=k_1]<\Pr[Y_1=1|\sum_{i=1}^kY_i=k_2] \label{eqn::simpleProbabilityExercise}
\end{align}
where each $Y_i$ is an independent Bernoulli random variable of probability $p_i\in(0,1)$ and $0\le k_1<k_2\le k$.  Furthermore, we can without loss of generality assume that $k_2=k_1+1$.

If $k_1=0$, then (\ref{eqn::simpleProbabilityExercise}) is clearly true, since the LHS is 0 while the RHS is non-zero.  So assume that $k_1>0$ and we can rewrite (\ref{eqn::simpleProbabilityExercise}) as follows:
\small
\begin{align*}
\frac{\Pr[Y_1=1\cap\sum_{i=1}^kY_i=k_1]}{\Pr[\sum_{i=1}^kY_i=k_1]} &<\frac{\Pr[Y_1=1\cap\sum_{i=1}^kY_i=k_1+1]}{\Pr[\sum_{i=1}^kY_i=k_1+1]} \\
\frac{p_1\Pr[\sum_{i=2}^kY_i=k_1-1]}{p_1\Pr[\sum_{i=2}^kY_i=k_1-1]+(1-p_1)\Pr[\sum_{i=2}^kY_i=k_1]} &<\frac{p_1\Pr[\sum_{i=2}^kY_i=k_1]}{p_1\Pr[\sum_{i=2}^kY_i=k_1]+(1-p_1)\Pr[\sum_{i=2}^kY_i=k_1+1]} \\
\left(1+\frac{(1-p_1)}{p_1}\cdot\frac{\Pr[\sum_{i=2}^kY_i=k_1]}{\Pr[\sum_{i=2}^kY_i=k_1-1]}\right)^{-1} &<\left(1+\frac{(1-p_1)}{p_1}\cdot\frac{\Pr[\sum_{i=2}^kY_i=k_1+1]}{\Pr[\sum_{i=2}^kY_i=k_1]}\right)^{-1}
\end{align*}
\normalsize
Therefore, it suffices to prove that:
\begin{align}
\frac{\Pr[\sum_{i=2}^kY_i=k_1]}{\Pr[\sum_{i=2}^kY_i=k_1-1]} &>\frac{\Pr[\sum_{i=2}^kY_i=k_1+1]}{\Pr[\sum_{i=2}^kY_i=k_1]} \nonumber \\
\Pr[\sum_{i=2}^kY_i=k_1]^2 &>\Pr[\sum_{i=2}^kY_i=k_1+1]\Pr[\sum_{i=2}^kY_i=k_1-1] \nonumber \\
\left(\sum_{S\subseteq\{2,\ldots,k\}:|S|=k_1}\prod_{i\in S}p_i\prod_{i\notin S}(1-p_i)\right)^2 &>\left(\sum_{S:|S|=k_1+1}\prod_{i\in S}p_i\prod_{i\notin S}(1-p_i)\right)\left(\sum_{S:|S|=k_1-1}\prod_{i\in S}p_i\prod_{i\notin S}(1-p_i)\right) \label{eqn::expansion}
\end{align}

After expanding, both sides are a sum of terms of the form
\begin{align}
\prod_{i=2}^kp_i^{a_i}(1-p_i)^{2-a_i} \label{eqn::2625}
\end{align}
where each $a_i$ is 0, 1, or 2 and the sum $\sum_{i=2}^ka_i$ equals $2k_1$, the total number of times that a ``positive'' term $p_i$ (as opposed to a ``negative'' term $(1-p_i)$) appears in the product.  Let $b$ denote the total number of $i=2,\ldots,k$ such that $a_i=1$, which must be even.  

Now, observe that the total number of times the term (\ref{eqn::2625}) appears in the LHS of the expansion of (\ref{eqn::expansion}) is $\binom{b}{b/2}$ (because we choose $b/2$ of the $b$ indices that are ``positive'' to come from the first bracket; the remaining $b/2$ must come from the second bracket) while the total number of times this term appears in the RHS is $\binom{b}{b/2+1}$ (because we choose $b/2+1$ of the $b$ indices that are ``positive'' to come from the first bracket), with the latter being strictly less.
Furthermore, none of these terms are 0, since all of the values of $p_i$ lie strictly between 0 and 1.  Therefore, the inequality is strict, completing the proof of the theorem.
\Halmos\endproof

\section{Alternate Clairvoyant Benchmark: the Deterministic Linear Program}\label{sec::dlp}

The deterministic linear program (DLP) relaxation \citep{GvR94,TvR06}, given valuation distributions $\vu_1,\ldots,\vu_T$, is defined as follows:
\begin{align}
\max\sum_{j=1}^mr^{(j)}\sum_{t=1}^Tx^{(j)}_t\Pr_{V_t\sim\vu_t}[V_t\ge r^{(j)}] & && \nonumber \\
\sum_{j=1}^m\sum_{t=1}^Tx^{(j)}_t\Pr_{V_t\sim\vu_t}[V_t\ge r^{(j)}] &\le k && \label{dlp::inv} \\
\sum_{j=1}^mx^{(j)}_t &\le1 && t\in[T] \nonumber \\
x^{(j)}_t &\ge0 && j\in[m],t\in[T] \nonumber
\end{align}
Let $\OPTLP(\vu_1,\ldots,\vu_T)$ denote the optimal objective value of the preceding LP.  It can be shown that $\OPTLP(\vu_1,\ldots,\vu_T)$ is also an upper bound on the expected revenue of the optimal dynamic programming policy knowing $\vu_1,\ldots,\vu_T$, since $x^{(j)}_t$ encapsulates the unconditional probability of the policy offering price $j$ to customer $t$.
Therefore, the competitive ratio relative to $\OPTLP(\vu_1,\ldots,\vu_T)$ also cannot be greater than $\CR(\cP)$.

In general, the value of $\OPTLP(\vu_1,\ldots,\vu_T)$ is incomparable to the value of our clairvoyant benchmark $\bE_{V_1\sim\vu_1,\ldots,V_T\sim\vu_T}[\OPT(V_1,\ldots,V_T)]$ from Section~\ref{sec::stochastic} (i.e. depending on the distributions $\vu_1,\ldots,\vu_T$, either could be strictly greater).
In Section~\ref{sec::stochastic}, we showed that the competitive ratio relative to $\bE_{V_1\sim\vu_1,\ldots,V_T\sim\vu_T}[\OPT(V_1,\ldots,V_T)]$ was still equal to $\CR(\cP)$, the same as the competitive ratio when the valuations were deterministic.
By contrast, we now show an example of a price set $\cP$ where the competitive ratio relative to $\OPTLP(\vu_1,\ldots,\vu_T)$ is strictly worse than $\CR(\cP)$.

Our example is based on the intuition that the DLP can exploit the fact that inventory constraint~\eqref{dlp::inv} only needs to hold in expectation.
That is, it is well-known that the DLP overestimates the optimum by a factor of $1-\frac{1}{e}$ even when the feasible price set $\cP$ consists of a singleton.  The example requires the starting inventory $k$ to be 1.  Without loss of generality assume $\cP=\{1\}$.  Consider $T$ customers, each of whom have a valuation exceeding 1 with probability $\frac{1}{T}$, and a valuation of 0 otherwise.  It is easy to check that $\OPTLP=1$ in this case, by setting $x^{(1)}_t=1$ for all $t\in[T]$.  Meanwhile, any algorithm cannot have expected revenue exceeding $1-(1-\frac{1}{T})^T$, where we have subtracted from 1 the probability of all customers having valuation 0.  As $T\to\infty$, $\dfrac{\bE[\ALG]}{\OPTLP}$ approaches $1-\dfrac{1}{e}$.

We now provide our example and result.
\begin{lemma}
Consider the model with independent, non-identical stochastic valuations, and let $\cP=\{1,r\}$, $k=1$.  For all $r\ge1$, there exists a distribution over sequences $(\vu_1,\ldots,\vu_T)$ such that for any online algorithm,
\begin{equation}\label{eqn::boundInPresentation}
\frac{\bE[\ALG(\vu_1,\ldots,\vu_T)]}{\bE[\OPTLP(\vu_1,\ldots,\vu_T)]}\le\min\{1-\frac{1}{e},\frac{r-r/e}{2r-1-r/e}\}.
\end{equation}
By Yao's minimax theorem, this shows that the competitive ratio relative to $\OPTLP(\vu_1,\ldots,\vu_T)$ cannot be greater than the RHS of~\eqref{eqn::boundInPresentation}.
\end{lemma}
If $r\le\frac{1}{1-1/e}$, then the RHS of (\ref{eqn::boundInPresentation}) is equal to $1-\frac{1}{e}\approx.632$.  However, if $r>\frac{1}{1-1/e}$, then we show that the upper bound is $\frac{r-r/e}{2r-1-r/e}$, which decreases to $\frac{e-1}{2e-1}\approx.387$ as $r\to\infty$.
Note that $\frac{r-r/e}{2r-1-r/e}$ is strictly less than $\frac{r}{2r-1}$ over $r\ge1$, which was our competitive ratio $\CR(\cP)$ established in Section~\ref{sec::stochastic} with $\cP=\{1,r\}$.

\proof{Proof.}
Suppose that $r>\frac{1}{1-1/e}$, and let $p=\frac{1}{r(1-1/e)}$, which is in $(0,1)$.  Consider the following distribution over $\vu_1,\ldots,\vu_T$:
\begin{itemize}
\item The first valuation distribution is deterministically $\vu_1=(v^{(0)}_1,v^{(1)}_1,v^{(2)}_1)=(0,1,0)$, i.e. the first customer deterministically has valuation 1.
\item With probability $p$, valuation distributions $\vu_2,\ldots,\vu_T$ are all equal to $(1-\frac{1}{T-1},0,\frac{1}{T-1})$.  When this occurs, each of the $T-1$ customers $2,\ldots,T$ are willing to pay $r$ with probability $\frac{1}{T-1}$, and 0 otherwise.
\item With probability $1-p$, valuation distributions $\vu_2,\ldots,\vu_T$ are all equal to $(1,0,0)$.  When this occurs, all customers $2,\ldots,T$ will never make a purchase.
\end{itemize}

We first compute the expected value of $\OPTLP(\vu_1,\ldots,\vu_T)$.  With probability $1-p$, $\OPTLP(\vu_1,\ldots,\vu_T)=1$, setting $x^{(1)}_1=1$.  With probability $p$, $\OPTLP(\vu_1,\ldots,\vu_T)=r$, setting $x^{(1)}_1=x^{(2)}_1=0$ and $x^{(2)}_2=\ldots=x^{(2)}_T=1$.  Therefore, $\bE[\OPTLP(\vu_1,\ldots,\vu_T)]=1-p+pr$.

We now consider the optimal strategy for the online algorithm.  It has to decide, at time 1, whether to sell the only unit of inventory at price 1, without knowing whether $\vu_2,\ldots,\vu_T$ are equal to $(1-\frac{1}{T-1},0,\frac{1}{T-1})$ or $(1,0,0)$.  Conditioned on it deciding to sell, $\ALG(\vu_1,\ldots,\vu_T)$ is deterministically 1.  Conditioned on it deciding to wait, $\ALG(\vu_1,\ldots,\vu_T)$ is $r$ with probability
\begin{equation}\label{eqn::2352}
\frac{1}{r(1-1/e)}\cdot(1-(1-\frac{1}{T-1})^{T-1}),
\end{equation}
and 0 otherwise.

We explain (\ref{eqn::2352}).  If the online algorithm decides to wait, then it will offer price $r$ to all customers beyond the first.  It gets a sale if $\vu_2=\ldots=\vu_T=(1-\frac{1}{T-1},0,\frac{1}{T-1})$, which occurs with probability $p=\frac{1}{r(1-1/e)}$, and further if at least 1 of the valuations $V_2,\ldots,V_T$ realizes to $r$, which yields the second term in (\ref{eqn::2352}).

Thus the expected revenue from deciding to wait is (\ref{eqn::2352}) multiplied by $r$, or
\begin{equation}\label{eqn::2151}
\frac{1}{(1-1/e)}\cdot(1-(1-\frac{1}{T-1})^{T-1}),
\end{equation}
which is always greater than 1.  Therefore, the online algorithm is better off waiting, in which case its expected revenue is (\ref{eqn::2151}).  Taking $T\to\infty$, (\ref{eqn::2151}) approaches 1.

As $T\to\infty$, the distribution we constructed over $\vu_1,\ldots,\vu_T$ is such that for the best online algorithm,
\begin{align*}
\frac{\bE[\ALG(\vu_1,\ldots,\vu_T)]}{\bE[\OPTLP(\vu_1,\ldots,\vu_T)]} &=\frac{1}{1-p+pr} \\
&=\frac{r(1-1/e)}{r(1-1/e)-1+r} \\
&=\frac{r-r/e}{2r-1-r/e},
\end{align*}
as desired.
\Halmos\endproof

\section{Proofs from Section~\ref{sec::stochastic}}\label{sec::stochasticPf}

\proof{Proof of Lemma~\ref{lem::exp_states}.}
We proceed by induction on $t$.  (\ref{eqn::inventory}) is true for $t=0$, since $\Pr[I^{\Exp}_0=k]=\Pr[I^{\VK}_0=k]=1$.

Now suppose $t>0$ and that (\ref{eqn::inventory}) has been established for time $t-1$.  Then for every $k'$ such that $\Pr[I^{\Exp}_{t-1}=k']>0$, (\ref{eqn::pricing}) holds by definition.  Indeed, since $\Pr[I^{\VK}_{t-1}=k']=\Pr[I^{\Exp}_{t-1}=k']$ by the inductive hypothesis, $\Pr[I^{\VK}_{t-1}=k']>0$ for such $k'$.

We now consider (\ref{eqn::inventory}) for time $t$.  Note that $I^{\Exp}_t=I^{\Exp}_{t-1}-\bI(V_t\ge\pexp_t)$.  Therefore,
\begin{eqnarray}\label{eqn::nonmax_inv}
\Pr[I^{\Exp}_t=k']=\Pr[I^{\Exp}_{t-1}=k'+1\cap V_t\ge\pexp_t]+\Pr[I^{\Exp}_{t-1}=k'\cap V_t<\pexp_t]
\end{eqnarray}
for $k'\in\{0,\ldots,k-1\}$, while
\begin{equation}\label{eqn::max_inv}
\Pr[I^{\Exp}_t=k]=\Pr[I^{\Exp}_{t-1}=k\cap V_t<\pexp_t].
\end{equation}

Now, for any $k'\in\{0,\ldots,k\}$, if $\Pr[I^{\Exp}_{t-1}=k']>0$, then the following can be derived:
\begin{eqnarray}
& & \Pr[V_t\ge\pexp_t|I^{\Exp}_{t-1}=k'] \nonumber \\
& = & \sum_{j=1}^{m+1}\Pr[V_t\ge\pexp_t|\pexp_t=r^{(j)},I^{\Exp}_{t-1}=k']\Pr[\pexp_t=r^{(j)}|I^{\Exp}_{t-1}=k'] \nonumber \\
& = & \sum_{j=1}^{m+1}\Pr[V_t\ge r^{(j)}]\Pr[\pexp_t=r^{(j)}|I^{\Exp}_{t-1}=k'] \nonumber \\
& = & \sum_{j=1}^{m+1}\Pr[V_t\ge r^{(j)}]\Pr[\pvk_t=r^{(j)}|I^{\VK}_{t-1}=k'] \nonumber \\
& = & \Pr[V_t\ge\pvk_t|I^{\VK}_{t-1}=k']. \label{eqn::derivation}
\end{eqnarray}
In the second equality, we remove the conditioning on $I^{\Exp}_{t-1}=k'$, since the valuation $V_t$ is an independent random variable unaffected by any history.  The third equality follows because we have already established (\ref{eqn::pricing}) for time $t$.  The final equality also requires independence.

By the inductive hypothesis that (\ref{eqn::inventory}) holds for time $t-1$, $\Pr[I^{\Exp}_{t-1}=k']=\Pr[I^{\VK}_{t-1}=k']$.  If $\Pr[I^{\Exp}_{t-1}=k']$ is also non-zero, then the following can be derived using (\ref{eqn::derivation}):
\begin{eqnarray}
\Pr[V_t\ge\pexp_t|I^{\Exp}_{t-1}=k']\Pr[I^{\Exp}_{t-1}=k'] & = & \Pr[V_t\ge\pvk_t|I^{\VK}_{t-1}=k']\Pr[I^{\VK}_{t-1}=k'] \nonumber \\
\Pr[V_t\ge\pexp_t\cap I^{\Exp}_{t-1}=k'] & = & \Pr[V_t\ge\pvk_t\cap I^{\VK}_{t-1}=k']. \label{eqn::plus_one}
\end{eqnarray}
If instead $\Pr[I^{\Exp}_{t-1}=k']=\Pr[I^{\VK}_{t-1}=k']=0$, then $\Pr[V_t\ge\pexp_t\cap I^{\Exp}_{t-1}=k']\le\Pr[I^{\Exp}_{t-1}=k']=0$.  Similarly, $\Pr[V_t\ge\pvk_t\cap I^{\VK}_{t-1}=k']=0$, and therefore, (\ref{eqn::plus_one}) still holds.

We can analogously to (\ref{eqn::derivation}) and (\ref{eqn::plus_one}) derive for all $k'\in\{0,\ldots,k\}$ that
\begin{equation}\label{eqn::nonplus_one}
\Pr[V_t<\pexp_t\cap I^{\Exp}_{t-1}=k']=\Pr[V_t<\pvk_t\cap I^{\VK}_{t-1}=k'].
\end{equation}

We can substitute (\ref{eqn::plus_one}) and (\ref{eqn::nonplus_one}) into (\ref{eqn::nonmax_inv}) to see that
\begin{eqnarray*}
\Pr[I^{\Exp}_t=k'] & = & \Pr[I^{\VK}_{t-1}=k'+1\cap V_t\ge\pvk_t]+\Pr[I^{\VK}_{t-1}=k'\cap V_t<\pvk_t] \\
& = & \Pr[\sum_{t'=1}^{t-1}X^{\VK}_{t'}=k-k'-1\cap X^{\VK}_t=1]+\Pr[\sum_{t'=1}^{t-1}X^{\VK}_{t'}=k-k'\cap X^{\VK}_t=0] \\
& = & \Pr[I^{\VK}_t=k']
\end{eqnarray*}
for all $k'\in\{0,\ldots,k-1\}$.  We can similarly substitute (\ref{eqn::nonplus_one}) into (\ref{eqn::max_inv}) to see that $\Pr[I^{\Exp}_t=k]=\Pr[I^{\VK}_t=k]$.  This completes the induction and the proof of Lemma~\ref{lem::exp_states}.
\Halmos\endproof

\proof{Proof of Theorem~\ref{thm::exp}.}
The following is straight-forward to derive:
\begin{eqnarray*}
\bE[\ALG^{\Exp}] & = & \sum_{t=1}^T\bE[\pexp_t\cdot\bI(V_t\ge\pexp_t)] \\
& = & \sum_{t=1}^T\sum_{j=1}^{m+1}r^{(j)}\Pr[V_t\ge r^{(j)}]\Pr[\pexp_t=r^{(j)}] \\
& = & \sum_{t=1}^T\sum_{j=1}^{m+1}r^{(j)}\Pr[V_t\ge r^{(j)}]\sum_{k'=0}^k\Pr[\pexp_t=r^{(j)}|I^{\Exp}_{t-1}=k']\Pr[I^{\Exp}_{t-1}=k'] \\
& = & \sum_{t=1}^T\sum_{j=1}^{m+1}r^{(j)}\Pr[V_t\ge r^{(j)}]\sum_{k'=0}^k\Pr[\pvk_T=r^{(j)}|I^{\VK}_{t-1}=k']\Pr[I^{\VK}_{t-1}=k'] \\
& = & \sum_{t=1}^T\sum_{j=1}^{m+1}r^{(j)}\Pr[V_t\ge r^{(j)}]\Pr[\pvk_t=r^{(j)}] \\
& = & \sum_{t=1}^T\bE[\pvk_t\cdot\bI(V_t\ge\pvk_t)].
\end{eqnarray*}
The second and sixth equalities use the independence of $V_t$, while the fourth equality uses both statements of Lemma~\ref{lem::exp_states}.  The final expression is equal to $\bE[\ALG^{\VK}]$, completing the proof of Theorem~\ref{thm::exp}.
\Halmos\endproof

\proof{Proof of Lemma~\ref{lem::coupling}.}
The first statement is easy to see.  Since every sample path fails at time $T+1$ by definition, for any sample path $\vH^{\Samp}_T$, it must have a unique \emph{first} point of failure in $[T+1]$, say $t'$.  $\vH^{\Samp}_T$ then falls under exactly one of the events, namely the one with $t=t'$ and $\vh_t=(0,\psamp_1,\vsamp_1,\ldots,0,\psamp_{t-1},\vsamp_{t-1},1)$.  Therefore, the events are mutually exclusive and collectively exhaustive.  The case for $\Exp$ is argued analogously.

The final statement is argued inductively.  For all $t\in\{0,\ldots,T\}$, let $\vg_t=(f_1,p_1,u_1,\ldots,f_t,p_t,u_t)$ be a vector of realizations to the end of time $t$, and let $\cG_t$ denote the set of such vectors containing no failures.  Let $\vG^{\Samp}_t=(\fsamp_1,\psamp_1,\vsamp_1,\ldots,\fsamp_t,\psamp_t,\vsamp_t)$, and

\noindent $\vG^{\Exp}_t=(\fexp_1,\pexp_1,\vexp_1,\ldots,\fexp_t,\pexp_t,\vexp_t)$.

We would like to inductively establish that $\Pr[\vG^{\Samp}_t=\vg_t]=\Pr[\vG^{\Exp}_t=\vg_t]$ for all $t\in\{0,\ldots,T\}$ and $\vg_t\in\cG_t$.  This is clearly true for $t=0$.  For $t>0$, take any $\vg_t\in\cG_t$, and we can write
\begin{align*}
\Pr[\vG^{\Samp}_t=\vg_t] &=\Pr[\vG^{\Samp}_{t-1}=\vg_{t-1}]\cdot\Pr[\fsamp_t=0|\vG^{\Samp}_{t-1}=\vg_{t=1}]\cdot\Pr[\psamp_t=p_t|\vG^{\Samp}_{t-1}=\vg_{t=1},\fsamp_t=0] \\
&\cdot\Pr[\vsamp_t=u_t|\vG^{\Samp}_{t-1}=\vg_{t=1},\fsamp_t=0,\psamp_t=p_t]; \\
\Pr[\vG^{\exp}_t=\vg_t] &=\Pr[\vG^{\exp}_{t-1}=\vg_{t-1}]\cdot\Pr[\fexp_t=0|\vG^{\exp}_{t-1}=\vg_{t=1}]\cdot\Pr[\pexp_t=p_t|\vG^{\exp}_{t-1}=\vg_{t=1},\fexp_t=0] \\
&\cdot\Pr[\vexp_t=u_t|\vG^{\exp}_{t-1}=\vg_{t=1},\fexp_t=0,\pexp_t=p_t].
\end{align*}
We will prove that $\Pr[\vG^{\Samp}_t=\vg_t]=\Pr[\vG^{\Exp}_t=\vg_t]$ by arguing that each term in the expression for $\Pr[\vG^{\Samp}_t=\vg_t]$ equals the corresponding term in the expression for $\Pr[\vG^{\Exp}_t=\vg_t]$.  The first terms are equal because of the inductive hypothesis.  The second terms are equal because both algorithms are sampling runs of Algorithm~\ref{alg::valuationTrackingProc} and trying to hit a run with $I^{\VK}_{t-1}=k-\sum_{t'=1}^{t-1}\bI(u_{t'}\ge p_{t'})$.  The third terms are identical because because we have conditioned on $\fsamp_t=0$.  The fourth terms are equal because $\vsamp_t$ and $\vexp_t$ are IID and none of the conditioning has any effect.

Having established this, note that for every $t\in[T+1]$ and $\vh_t\in\cF_t$ there exists a unique $\vg_{t-1}\in\cG_{t-1}$ such that $\vg_{t-1}$ is a prefix of $\vh_t$.  We know that for this $\vg_{t-1}$, $\Pr[\vG^{\Samp}_{t-1}=\vg_{t-1}]=\Pr[\vG^{\Exp}_{t-1}=\vg_{t-1}]$.  Therefore, it suffices to prove that $\Pr[\fsamp_t=0|\vG^{\Samp}_{t-1}=\vg_{t-1}]=\Pr[\fexp_t=0|\vG^{\Exp}_{t-1}=\vg_{t-1}]$.  By the same argument as the previous paragraph, these two probabilities are equal.  Therefore, $\Pr[\vH^{\Samp}=\vh_t]=\Pr[\vH^{\Exp}=\vh_t]$, completing the proof of Lemma~\ref{lem::coupling}.
\Halmos\endproof

\proof{Proof of Theorem~\ref{thm::poly}.}
Applying Lemma~\ref{lem::prob_failing} to (\ref{eqn::final_frontier}), we see that
\begin{eqnarray*}
\bE[\ALG^{\Samp}] & \ge & \bE[\ALG^{\Exp}]-\bE[\OPT]\Big(\sum_{t=1}^T\frac{1}{eCt^2}\Big) \\
& \ge & \bE[\ALG^{\Exp}]-\bE[\OPT]\frac{1}{e\lceil\frac{6}{e\pi^2\ve}\rceil}\Big(\frac{\pi^2}{6}\Big) \\
& \ge & \bE[\ALG^{\Exp}]-\ve\cdot\bE[\OPT]
\end{eqnarray*}
Furthermore, we know from Theorem~\ref{thm::exp} that $\bE[\ALG^{\Exp}]=\bE[\ALG^{\VK}]=\frac{1}{q}\bE[\OPT]$.  This establishes the competitiveness.

The statement about runtime also follows easily from the specification of Algorithm~\ref{alg::sampling} since the number of sample runs during each time period $t$, $\lceil\frac{6}{e\pi^2\ve}\rceil(k+1)t^2$, is polynomial in $\frac{1}{\ve}$.
\Halmos\endproof

\proof{Proof.}
Consider any $t\in[T]$ and $\vh_t\in\cF_t$.  We have
\begin{eqnarray}
\bE[\ALG^{\Samp}|\vH^{\Samp}_t=\vh_t] & \ge & \bE\Big[\sum_{t'=1}^{t-1}\psamp_{t'}\cdot\bI(\vsamp_{t'}\ge\psamp_{t'})\Big|\vH^{\Samp}_t=\vh_t\Big] \nonumber \\
& = & \sum_{t'=1}^{t-1}p_{t'}\cdot\bI(u_{t'}\ge p_{t'}). \label{eqn::4984}
\end{eqnarray}

Meanwhile, $\bE[\ALG^{\Exp}|\vH^{\Exp}_t=\vh_t]$ can be decomposed into
\begin{equation}\label{eqn::decompose}
\sum_{t'=1}^{t-1}p_{t'}\cdot\bI(u_{t'}\ge p_{t'})+\bE\Big[\sum_{t'=t}^T\pexp_{t'}\cdot\bI(\vexp_{t'}\ge\pexp_{t'})\Big|\vH^{\Exp}_t=\vh_t\Big].
\end{equation}
We elaborate on the second term in (\ref{eqn::decompose}).  Clearly, $\sum_{t'=t}^T\pexp_t\cdot\bI(\vexp_t\ge\pexp_t)$ cannot exceed the sum of the $\min\{k,T-t+1\}$ largest valuations to appear during $t,\ldots,T$, which we denote by $\OPT(\vexp_t,\ldots,\vexp_T)$.  Furthermore, the random valuations $\vexp_t,\ldots,\vexp_T$ are independent of the history $\vH^{\Exp}_t$ up to time $t$, so we can remove the conditioning and upper-bound (\ref{eqn::decompose}) with
\begin{equation}\label{eqn::1919}
\sum_{t'=1}^{t-1}p_{t'}\cdot\bI(u_{t'}\ge p_{t'})+\bE[\OPT(\vexp_t,\ldots,\vexp_T)].
\end{equation}
The expectation in (\ref{eqn::1919}) is with respect to $\vexp_t,\ldots,\vexp_T$ being drawn independently according to $\vu_t,\ldots,\vu_T$.  (\ref{eqn::1919}) in turn is no greater than $\sum_{t'=1}^{t-1}p_{t'}\cdot\bI(u_{t'}\ge p_{t'})+\bE[\OPT(\vexp_1,\ldots,\vexp_T)]$, where the random variables $\vexp_1,\ldots,\vexp_{t-1}$ are \emph{not} conditioned on the event $\vH^{\Exp}_t=\vh_t$.  The proof of the lemma concludes by comparing this expression with (\ref{eqn::4984}).
\Halmos\endproof

\proof{Proof.}
Consider any $t\in[T]$.  For all $\vh_t\in\cF_t$, let $G(\vh_t)=(f_1,p_1,u_1,\ldots,f_{t-1},p_{t-1},u_{t-1})$, which is the vector of the first $3(t-1)$ entries in $\vh_t$.  Let $\vG^{\Exp}_{t-1}=(\fexp_1,\pexp_1,\vexp_1,\ldots,\fexp_{t-1},\pexp_{t-1},\vexp_{t-1})$, which is a vector of $3(t-1)$ random variables.

We can write $\sum_{\vh_t\in\cF_t}\Pr[\vH^{\Exp}_t=\vh_t]$ as
\begin{equation}\label{eqn::1591}
\sum_{\vh_t\in\cF_t}\Pr[\fexp_t=1|\vG^{\Exp}_{t-1}=G(\vh_t)]\Pr[\vG^{\Exp}_{t-1}=G(\vh_t)].
\end{equation}

Now, for each $\vh_t\in\cF_t$, $\Pr[\fexp_t=1|\vG^{\Exp}_{t-1}=G(\vh_t)]$ is the probability that all $C(k+1)t^2$ independent runs of Algorithm~\ref{alg::valuationTrackingProc} fail to match the inventory remaining at the start of time $t$ according to $\vh_t$.  For convenience, define $I(\vh_t)=k-\sum_{t'=1}^{t-1}\bI(u_{t'}\ge p_{t'})$.  Then
\begin{equation}\label{eqn::9456}
\Pr[\fexp_t=1|\vG^{\Exp}_{t-1}=G(\vh_t)]=(1-\Pr[I^{\VK}_{t-1}=I(\vh_t)])^{C(k+1)t^2},
\end{equation}
where $I^{\VK}_{t-1}$ is the total inventory remaining at the start of time $t$ in a run of Algorithm~\ref{alg::valuationTrackingProc}.
%, as defined in Definition~\ref{def::diff_algs}.

Therefore, we can partition the $\vh_t$ in $\cF_t$ by $I(\vh_t)$.  For all $k'\in\{0,\ldots,k\}$, define $\rho_{t,k'}=\Pr[I^{\VK}_{t-1}=k']$.  The following can be derived by substituting (\ref{eqn::9456}) into (\ref{eqn::1591}):
\begin{eqnarray}
\sum_{\vh_t\in\cF_t}\Pr[\vH^{\Exp}_t=\vh_t] & = & \sum_{k'=0}^k(1-\rho_{t,k'})^{C(k+1)t^2}\sum_{\vh_t\in\cF_t:I(\vh_t)=k'}\Pr[\vG^{\Exp}_{t-1}=G(\vh_t)] \nonumber \\
& \le & \sum_{k'=0}^k\exp(-\rho_{t,k'}C(k+1)t^2)\sum_{\vh_t\in\cF_t:I(\vh_t)=k'}\Pr[\vG^{\Exp}_{t-1}=G(\vh_t)]. \label{eqn::9847}
\end{eqnarray}

At this point, we would like to argue that $\sum_{\vh_t\in\cF_t:I(\vh_t)=k'}\Pr[\vG^{\Exp}_{t-1}=G(\vh_t)]\le\Pr[I^{\Exp}_{t-1}=k']$.  To see this, note that $\sum_{\vh_t\in\cF_t:I(\vh_t)=k'}\Pr[\vG^{\Exp}_{t-1}=G(\vh_t)]=\Pr[I^{\Exp}_{t-1}=k'\cap(\fexp_1=\ldots=\fexp_{t-1}=0)]$.
%we are only summing the probabilities of sequences $G(\vh_t)=(F_1,P_1,V_1,\ldots,F_{t-1},P_{t-1},V_{t-1})$ such that $F_1=\ldots=F_{t-1}=0$.  We are omitting the sequences where $k-\sum_{t'=1}^{t-1}\bI(V_{t'}\ge P_{t'})=I_{t-1}=k'$ but some failure indicator $F_{t'}$, $t'\in[t-1]$, is true.  In other words, , which is no greater than $\Pr[I_{t-1}=k']$.

Applying the second statement of Lemma~\ref{lem::exp_states}, we see that $\Pr[I^{\Exp}_{t-1}=k']=\rho_{t,k'}$.  Substituting into (\ref{eqn::9847}), the following can be derived:
\begin{eqnarray*}
\sum_{\vh_t\in\cF_t}\Pr[\vH^{\Exp}_t=\vh_t] & \le & \sum_{k'=0}^k\rho_{t,k'}\exp(-\rho_{t,k'}C(k+1)t^2) \\
& \le & \sum_{k'=0}^k\frac{1}{C(k+1)t^2}\exp(-1) \\
& = & \frac{1}{eCt^2}.
\end{eqnarray*}
The second inequality holds because for a single $\rho_{t,k'}\in[0,1]$, the function $\rho_{t,k'}e^{-\rho_{t,k'}C(k+1)t^2}$ is maximized at $\rho_{t,k'}=\frac{1}{C(k+1)t^2}$.  The proof of the lemma is now complete.
\Halmos\endproof

\section{Proofs from Section~\ref{sec::extensions}}\label{sec::extensionsPf}

\proof{Proof of Lemma~\ref{lem::contPrice1}.}
We proceed by induction on $t$.  (\ref{eqn::cont_depletion}) is true at time $t=0$, where $\bE[S_{i,0}]=0$ and $w_{i,0}=0$ for all $i\in[k]$.

Now suppose we are at the end of some time $t>0$ and (\ref{eqn::cont_depletion}) was true at the end of time $t-1$.  It suffices to prove that $\bE[S_{i_t,t}]=\bI(w_{i_t,t}>0)\cdot\frac{1+\ln w_{i_t,t}}{1+\ln R}$.  This is immediate if $V_t\le w_{i_t,t-1}$, by the induction hypothesis.  Otherwise, if $V_t>w_{i_t,t-1}$, we consider two cases.  Let $v=w_{i_t,t-1}$ for brevity.

We know that $\bE[S_{i_t,t}]=\bE[S_{i_t,t-1}]+\bE[X_t|S_{i_t,t-1}=0]\cdot\Pr[S_{i_t,t-1}=0]$.

If $v=0$, then this equals 
\begin{eqnarray*}
\Pr[X_t=1|S_{i_t,t-1}=0] & = & \frac{1}{1+\ln R}(1+\int_1^{w_{i,t}}\frac{1}{r}dr) \\
& = & \frac{1+\ln w_{i,t}}{1+\ln R}
\end{eqnarray*}
as desired.  On the other hand, if $v>0$, then
\begin{eqnarray*}
\bE[S_{i_t,t}] & = & \frac{1+\ln v}{1+\ln R}+\frac{1}{\ln R-\ln v}\int_v^{w_{i,t}}\frac{1}{r}dr\Big(1-\frac{1+\ln v}{1+\ln R}\Big) \\
& = & \frac{1+\ln v}{1+\ln R}+\frac{\ln w_{i,t}-\ln v}{\ln R-\ln v}\Big(\frac{\ln R-\ln v}{1+\ln R}\Big) \\
& = & \frac{1+\ln w_{i,t}}{1+\ln R}.
\end{eqnarray*}

This completes the induction and the proof of the lemma.
\Halmos\endproof

\proof{Proof of Lemma~\ref{lem::contPrice2}.}
Let $t\in[T]$ be any time step for which $V_t>w_{i_t,t-1}$.  Again, let $v$ denote $\ell_{i_t,t-1}$, and we consider the two cases $v=0$ and $v>0$.  If $v=0$, then
\begin{eqnarray*}
\bE[P_tX_t] & = & \frac{1}{1+\ln R}\Big(1+\int_1^Rr\bE[X_t|P_t=r]\frac{1}{r}dr\Big) \\
& = & \frac{1}{1+\ln R}\Big(1+\int_1^R\bI[w_{i_t,t}\ge r]dr\Big) \\
& = & \frac{1}{1+\ln R}(1+w_{i_t,t}-1).
\end{eqnarray*}
In the first equality, a sale is guaranteed if $P_t=1$, earning revenue 1.  The final term is the desired expression.

If $v>0$, then
\begin{eqnarray*}
\bE[P_tX_t] & = & \frac{1}{\ln R-\ln v}\Big(\int_v^Rr\bE[X_t|P_t=r]\frac{1}{r}dr\Big)\Big(\frac{\ln R-\ln v}{1+\ln R}\Big) \\
& = & \frac{1}{1+\ln R}\Big(\int_v^R\bI[w_{i_t,t}\ge r]dr\Big) \\
& = & \frac{1}{1+\ln R}(w_{i_t,t}-v)
\end{eqnarray*}
as desired.
\Halmos\endproof

\proof{Proof of Proposition~\ref{prop::no_info}.}
Let the starting inventory $k=1$.

First, it is easy to see that if the distribution of $P_1$ is not such that $\Pr[P_1=r^{(j)}]=\frac{q^{(j)}}{q}$ for all $j\in[m]$, then for some deterministic instance consisting of a single valuation in $\{r^{(1)},\ldots,r^{(m)}\}$, $\frac{\bE[\ALG]}{\OPT}$ will be strictly less than $\frac{1}{q}$.  Therefore we can without loss of generality assume that $\Pr[P_1=r^{(j)}]=\frac{q^{(j)}}{q}$ for all $j\in[m]$ (regardless of whether valuations can be 0).

Now suppose $m\ge3$.  Consider the distribution of $P_2$ conditioned on $X_1=0$.  If $\Pr[P_2\ge r^{(m)}|X_1=0]=1$, then consider the instance $T=2,V_1=1,V_2=r^{(m-1)}$.  $\OPT=r^{(m-1)}$, which exceeds 1, since $m\ge3$.  Meanwhile, $q\bE[\ALG]=q\frac{1}{q}<\OPT$.  On the other hand, if $\Pr[P_2\ge r^{(m)}|X_1=0]<1$, then consider the instance $T=3,V_1=V_2=r^{(m-1)},V_3=r^{(m)}$.  $\OPT=r^{(m)}$.  $q\bE[P_1X_1]=r^{(m-1)}$, while $\bE[X_1]=1-\frac{q^{(m)}}{q}$.  The best case for the algorithm, given that $V_2=r^{(m-1)}$, is $P_2=r^{(m-1)}$ when $P_2<r^{(m)}$.  Let $\Pr[P_2=r^{(m-1)}|X_1=0]=\alpha$, which we know is positive.  In this case, $q\bE[P_2X_2]=r^{(m-1)}\alpha q^{(m)}$ and $\bE[X_2]=\alpha\frac{q^{(m)}}{q}$.  Hence $q\bE[P_3X_3]$ is at most $qr^{(m)}(1-\bE[X_1+X_2])=r^{(m)}(1-\alpha)q^{(m)}$.  All in all, $q\bE[\ALG]$ is at most
\begin{eqnarray*}
r^{(m-1)}+r^{(m-1)}\alpha(1-\frac{r^{(m-1)}}{r^{(m)}})+r^{(m)}(1-\alpha)(1-\frac{r^{(m-1)}}{r^{(m)}}) & = & r^{(m)}+\alpha(2r^{(m-1)}-\frac{(r^{(m-1)})^2}{r^{(m)}}-r^{(m)}) \\
& = & r^{(m)}-\frac{\alpha}{r^{(m)}}(r^{(m)}-r^{(m-1)})^2
\end{eqnarray*}
The term getting subtracted is non-zero since $\alpha>0$ and $r^{(m)}>r^{(m-1)}$.  Therefore, $q\bE[\ALG]<\OPT$.  This completes the proof when $m\ge3$, since $\CR(\cP)=1/q$.

The case where $m=2$ and valuations can be 0 is argued analogously.  If $\Pr[P_2\ge r^{(2)}]=1$, then consider the instance $T=2,V_1=0,V_2=r^{(1)}$.  If $\Pr[P_2<r^{(2)}]=1$, then consider the instance $T=3,V_1=V_2=r^{(1)},V_3=r^{(2)}$.  In both cases, it can be seen that $q\bE[\ALG]<\OPT$, completing the proof of Proposition~\ref{prop::no_info}.
\Halmos\endproof

\proof{Proof of Proposition~\ref{prop::random_yes}.}
Consider any realization of the valuations, $V_1,\ldots,V_T$.  Iteratively define the following quantities, for $j$ from $m$ down to 1:
\begin{equation}
n^{(j)}=\min\Big\{\sum_{t=1}^T\bI(V_t=r^{(j)}),k-\sum_{j'=j+1}^mn^{(j')}\Big\}.
\end{equation}
Essentially, for each $j$, $n^{(j)}$ denotes the number of valuations equal to $r^{(j)}$ that should be picked out when picking out the $\min\{k,T\}$ largest valuations.  $\OPT$ is then equal to $\sum_{j=1}^mr^{(j)}n^{(j)}$.

Now consider the execution of the policy on this instance.  For all $j\in[m]$, if the random fixed price $P$ is equal to $r^{(j)}$, then the number of sales will be equal to $\min\{\sum_{t=1}^T\bI(V_t\ge r^{(j)}),k\}$, which by definition is equal to $\sum_{j'=j}^mn^{(j')}$.  Therefore,
\begin{eqnarray*}
\bE[\ALG] & = & \frac{1}{q}\sum_{j=1}^mq^{(j)}r^{(j)}\sum_{j'=j}^mn^{(j')} \\
& = & \frac{1}{q}\sum_{j'=1}^mn^{(j')}\sum_{j=1}^{j'}\big(1-\frac{r^{(j-1)}}{r^{(j)}}\big)r^{(j)} \\
& = & \frac{1}{q}\sum_{j'=1}^mn^{(j')}r^{(j')}
\end{eqnarray*}
which equals $\frac{1}{q}\OPT$, completing the proof that the random fixed price is $\CR(\cP)$-competitive.
\Halmos\endproof

\end{APPENDICES}

%%%%%%%%%%%%%%%%%
\end{document}